\documentclass[twocolumn,preprintnumbers]{revtex4}

\usepackage{graphicx}
\usepackage{epsf}
\usepackage{amsmath,amssymb}
\newcommand{\bvec}{\boldsymbol}

\begin{document}
%\preprint{KUNS-2503}
\title{Excitation energy shift and size difference of low-energy levels in 
$p$-shell $\Lambda$ hypernuclei
% $^{7}_\Lambda \textrm{Li}$,  $^{9}_\Lambda \textrm{Be}$
%$^{11}_\Lambda \textrm{B}$, $^{12}_\Lambda \textrm{B,C}$,
%and $^{13}_\Lambda \textrm{C}$
}
\author{Yoshiko Kanada-En'yo}
\affiliation{Department of Physics, Kyoto University, Kyoto 606-8502, Japan}
\begin{abstract}
We investigated structures of low-lying $0s$-orbit $\Lambda$ states 
in $p$-shell  $\Lambda$ hypernuclei ($^A_\Lambda Z$) by applying 
microscopic cluster models for nuclear structure and a single-channel
 folding potential model for a $\Lambda$ particle. 
For $A>10$ systems, the size reduction of core nuclei is small,
and the core polarization effect is regarded as a higher-order perturbation in the $\Lambda$ binding.
The present calculation qualitatively describes the systematic trend of experimental data
for excitation energy change from $^{A-1} Z$ to $^A_\Lambda Z$, in 
$A>10$ systems.
The energy change shows a clear correlation with the 
nuclear size difference between the ground and excited states.
In $^7_\Lambda \textrm{Li}$ and $^9_\Lambda \textrm{Be}$,
the significant shrinkage of cluster structures occurs consistently with 
the prediction of other calculations. 
%The results reproduce well the experimental
%$E2$ transition strengths in $^6\textrm{Li}$ and $^7_\Lambda \textrm{Li}$ without using effective charges. 
\end{abstract}
\maketitle

\section{Introduction}
Owing to high-resolution $\gamma$-ray measurement experiments, spectra of low-lying states 
of various $p$-shell $\Lambda$ hypernuclei have been revealed in these years \cite{Hashimoto:2006aw,Tamura:2010zz,Tamura:2013lwa}.
Measured energy spectra and electromagnectic transitions are useful information to know properties of 
$\Lambda$-nucleon($N$) interactions and also helpful to investigate impurity effects of a $\Lambda$ particle on nuclear systems. 
In order to theoretically study structures of $p$-shell $\Lambda$ hypernuclei,  various calculations have been performed with 
cluster models \cite{Motoba:1984ri,motoba85,Yamada:1985qr,Yu:1986ip,Hiyama:1996gv,Hiyama:1997ub,Hiyama:1999me,Hiyama:2000jd,Hiyama:2002yj,Hiyama:2006xv,Hiyama:2010zzc,Cravo:2002jv,Suslov:2004ed,Mohammad:2009zza,Zhang:2012zzg,Funaki:2014fba,Funaki:2017asz}, shell models \cite{Gal:1971gb,Gal:1972gd,Gal:1978jt,Millener:2008zz,Millener:2010zz,Millener:2012zz}, mean-field and beyond mean-field models \cite{Guleria:2011kk,Vidana:2001rm,Zhou:2007zze,Win:2008vw,Win:2010tq,Lu:2011wy,Mei:2014hya,Mei:2015pca,Mei:2016lce,Schulze:2014oia}, hyper antisymmetrized molecular dynamics (HAMD) model
\cite{Isaka:2011kz,Isaka:2015xda,Homma:2015kia,Isaka:2016apm,Isaka:2017nuc},  and nore-core shell model \cite{Wirth:2014apa}, and so on.

Since a $\Lambda$ particle is free from Pauli blocking and $\Lambda$-$N$ interactions are weaker than $N$-$N$ interactions, 
the $\Lambda$ spin degree of freedom in $\Lambda$ hypernuclei more or less weakly couples with core nuclei in general. 
Therefore, the $\Lambda$ particle in $\Lambda$ hypernuclei 
can be regarded as an impurity of the nuclear system. Indeed, there are many theoretical works discussing 
$\Lambda$ impurity effects of on nuclear structures such as 
shrinkage effects on cluster structures \cite{Motoba:1984ri,motoba85,Yu:1986ip,Hiyama:1996gv,Hiyama:1997ub,Hiyama:1999me,Hiyama:2002yj,Hiyama:2006xv,Hiyama:2010zzc,Isaka:2015xda,Homma:2015kia,Isaka:2011zza,Sakuda:1987pk,Yamada:1984ii} and effects on nuclear deformations \cite{Zhou:2007zze,Win:2008vw, Win:2010tq,Lu:2011wy,Isaka:2011aw,Isaka:2011kz,Isaka:2016apm,Lu:2014wta}.
One of the famous phenomena is the shrinkage of $^7_\Lambda\textrm{Li}$, which has been theoretically predicted \cite{Motoba:1984ri,motoba85}
and later evidenced  experimentally through the $E2$ transition strength measurement \cite{Tanida:2000zs}. 
The dynamical effects of $\Lambda$ on nuclear structures can be significant in the case that  
core nuclei are fragile systems such as weakly bound systems and shape softness (or coexistence) ones.
However, except for such cases, 
dynamical change of nuclear structure (the core polarization) is expected to be minor in general
because of the weaker $\Lambda$-$N$ interactions and no Pauli blocking.
In this context, there might be a chance to probe original properties of core nuclear structures by a $\Lambda$ particle 
perturvatively appended to the nuclear system. 

Let us focus on energy spectra of $p$-shell $\Lambda$ hypernuclei.
% because detailed experimental information is being increasing rapidly. 
The low-energy levels are understood as core excited states with  
a  $0s$-orbit $\Lambda$ ($(0s)_\Lambda$ states). When we consider the $\Lambda$ particle as an impurity giving a perturbation to the core nuclear system, 
the first-order perturbation on the energy spectra, that is, change of excitation energies by the $\Lambda$ particle, 
comes from structure difference between the ground and excited states through the $\Lambda$-$N$ interactions, 
whereas dynamical structure change gives second-order perturbation effects on the energy spectra.
For excited states with structures much different 
from that of the ground state,  the $\Lambda$ particle can give significant effect on 
energy spectra as discussed by Isaka {\it et al.} for Be isotopes \cite{Isaka:2015xda,Homma:2015kia}. 
In this concern, it is meaningful to look at  excitation energy shifts, that is, excitation energy changes from $^{A-1}Z$ to 
$^A_\Lambda Z$,  in  available data.
For simplicity, we here ignore the $\Lambda$  intrinsic spin degree of freedom because spin dependence of the 
$\Lambda$-$N$ interactions is weak.
In the observed energy spectra of $^{10}\textrm{B}$-$^{11}_\Lambda\textrm{B}$, 
$^{11}\textrm{B}$-$^{12}_\Lambda\textrm{B}$, $^{11}\textrm{C}$-$^{12}_\Lambda\textrm{C}$,  and 
$^{12}\textrm{C}$-$^{13}_\Lambda\textrm{C}$ systems, one can see that the excitation energies $(E_x)$  for 
$^{10}\textrm{B}(3^+_1)$, $^{11}\textrm{B}(1/2^-_1,3/2^-_2)$,  $^{11}\textrm{C}(1/2^-_1,3/2^-_2)$
and $^{12}\textrm{C}(2^+_1)$ are significantly raised by the $\Lambda$ particle in $^A_\Lambda Z$ systems
compared with those in $^{A-1}Z$ systems.  On the other hand, the situation is opposite
in $^6\textrm{Li}$-$^{7}_\Lambda\textrm{Li}$ systems. 
the $E_x(3^+)$ is decreased by the $\Lambda$ particle. 
To systematically comprehend the energy spectra of $p$-shell $\Lambda$ hypernuclei, it is worth to
examine  the excitation energy shifts and their link with the structure difference between  
the ground and excited states. 

Precise data of spectroscopy in various $\Lambda$ hypernuclei are becoming available and they provide
fascinating physics in nuclear many-body systems consisting of protons, neutrons, and $\Lambda$s.
Sophisticated calculations have been achieved mainly in light nuclei and greatly contributed to 
progress of physics of hypernuclei.
Nevertheless,  systematic studies for energy spectra of hypernuclei in a wide mass-number region 
are still limited compared with those for ordinary nuclei, for which various structure models have been 
developed and used for intensive and extensive studies. It is time to extend application of such 
structure models developed for ordinary nuclei to hypernuclei. To this end, it might be helpful 
to propose  a handy and economical treatment of a $\Lambda$ particle and core polarization in $\Lambda$ hyper 
nuclei that can be applied to 
general structure models. 

Our first aim, in this paper, is to investigate energy spectra of low-lying $(0s)_\Lambda$ states 
 in $p$-shell  $\Lambda$ hypernuclei. 
A particular attention is paid to the excitation energy shifts by the $\Lambda$ and their link with structures of core nuclei. 
The second aim is that we are to propose a handy treatment of the $\Lambda$ particle in $\Lambda$ hypernuclei and to check its
phenomenological applicability. 
To describe detailed structures of the ground and excited states of core nuclei, 
we apply the generator coordinate method (GCM) \cite{GCM1,GCM2} of 
microscopic $\alpha+d$, $2\alpha$, and $2\alpha+d$ cluster models for $^{6}\textrm{Li}$,
$^{8}\textrm{Be}$, and $^{10}\textrm{B}$, respectively, 
and that of extended $2\alpha+t$ and $3\alpha$ cluster models with the cluster breaking
	for $^{11}\textrm{B}$, $^{11}\textrm{C}$, and $^{12}\textrm{C}$. 
For description of $(0s)_\Lambda$ states in $\Lambda$ hypernuclei, a single $S$-wave channel calculation with a folding potential model is performed.
Namely,  the $\Lambda$-nucleus potentials are
constructed by folding $\Lambda$-$N$ interactions with the nuclear density
calculated by the microscopic cluster models.
As a core polarization effect, 
the core size reduction is taken into account in a simple way.
%In the present model, difference in structures between the ground and excited states is reflected 
%in the $\Lambda$-nucleus potentials through the spherical nuclear density. 
%The core polarization can contribute to excitation energy shifts through density (size) change of the core nuclei. 

This paper is organized as follows. In the next section, we describe formalism of the present model.
The adopted effective $N$-$N$ and $\Lambda$-$N$ interactions are explained in Sec.~\ref{sec:interactions}.
The results are shown in Sec.~\ref{sec:results}, and discussions are given in Sec.~\ref{sec:discussions}. 
Finally, the paper is summarized in Sec.~\ref{sec:summary}.

\section{formulation}\label{sec:model}

\subsection{microscopic cluster model for core nuclei}
Structures of core nuclei are calculated by the microscopic cluster models
with the GCM using the Brink-Bloch cluster wave functions \cite{Brink66}.
In the cluster GCM calculations, we superpose the microscopic 
$\alpha+d$, $2\alpha$, and $2\alpha+d$, $2\alpha+t(h)$, and $3\alpha$
wave functions for $^6\textrm{Li}$, $^8\textrm{Be}$,  $^{10}\textrm{B}$, 
 $^{11}\textrm{B(C)}$, and  $^{12}\textrm{C}$.  

For a system consisting of $C_1, \ldots, C_k$ clusters ($k$ is the number of clusters),  
the Brink-Bloch cluster  wave function is given as 
\begin{eqnarray}
&&\Phi_\textrm{BB}(\bvec{S}_1,\ldots, \bvec{S}_k; \bvec{r}_1\sigma_1,\ldots \bvec{r}_A\sigma_A)\nonumber\\
&&={\cal A} \left[ \phi_{C_1}\left(\bvec{S}_1; \bvec{r}_1\sigma_1, \ldots, \bvec{r}_{A_1}\sigma_{A_1}\right) \right.\nonumber\\
&&\cdots \left. \phi_{C_k}\left(\bvec{S}_k;\bvec{r}_{A-A_{k}}\sigma_{A-A_{k}}, \ldots, \bvec{r}_{A_k}\sigma_{A_k}\right)
\right], 
\end{eqnarray}
where $\bvec{S}_j$ indicates the position parameter of the $C_j$ cluster,  $\bvec{r}_i$ and $\sigma_i$ indicate
the coordinate and spin-isospin configuration of the $i$th nucleon,  
 ${\cal A}$ is the antisymmetrizer of all nucleons, $A$ is the mass number, and $A_j$ is the mass number of the $C_j$ cluster.
The $A_j$-nucleon wave function 
$\phi_{C_j}$ for the $C_j$ cluster is written by
the $(0s)^{A_j}$ harmonic oscillator shell model wave function with the center shifted to the position $\bvec{S}_j$. 
The intrinsic spin configurations of $d$, $t(h)$, and $\alpha$ clusters are $S=1$, 1/2, and 0 states, respectively.
The width parameter $\nu= 1/(2b^2)$ ($b$ is the size parameter) of the harmonic oscillator
is set to be a common value so that the center of mass (cm) motion 
can be removed exactly. In the present work, we use the same 
parameter $\nu=0.235$ fm$^{-2}$ as that used in Ref.~\cite{Suhara:2014wua} which reasonably reproduces the ground-state sizes of $p$-shell nuclei.
The Brink-Bloch cluster wave function is a fully microscopic $A$-nucleon wave function, in which 
the degrees of freedom and antisymmetrization of $A$ nucleons are taken into account, 
differently from non-microscopic cluster models (simple $k$-body potential models)
and such semi-microscopic cluster models as the orthogonal condition model (OCM) \cite{Saito:1969zz}.  

To take into account inter-cluster motion, the GCM is performed with respect to the cluster center parameters 
$\bvec{S}_j$. Namely, the GCM wave function $\Psi(J^\pi_n)$ for the 
$J^\pi_n$ state is expressed by linear combination of the spin-parity projected 
Brink-Bloch wave functions with various configurations of $\bvec{S}_j$ as 
\begin{equation}\label{eq:gcm-wf}
\Psi(J^\pi_n)=\sum_{\bvec{S}_1,\ldots, \bvec{S}_k} \sum_{K} c^{J^\pi_n}_{\bvec{S}_1,\ldots, \bvec{S}_k,K}
P^{J\pi}_{MK} \Phi_\textrm{BB}(\bvec{S}_1,\ldots, \bvec{S}_k),
\end{equation}
where $P^{J\pi}_{MK}$ is the spin-parity projection operator. The coefficients 
$c^{J^\pi_n}_{\bvec{S}_1,\ldots, \bvec{S}_k,K}$ are determined by diagonalization of the Hamiltonian and norm matrices.
In the present calculation, 
for two-cluster systems of $\alpha+d$ and $2\alpha$, $\bvec{S}_k$ is chosen to be 
$\bvec{S}_1-\bvec{S}_2=(0,0,d)$ with $d=\{1,2,\cdots,15$ fm\}. 
For three-cluster systems of $2\alpha+d$, $2\alpha+t(h)$, $3\alpha$, $\bvec{S}_k$ is chosen to be 
\begin{eqnarray}
&&\bvec{S}_1-\bvec{S}_2=(0,0,d), \\
&&\bvec{S}_3-\frac{A_2\bvec{S}_1+A_1\bvec{S}_2}{A_1+A_2}
=(r\sin \theta,0,r \cos \theta),
\end{eqnarray}
with $d=\{1.2,2.4,\ldots,4.2$ fm\}, $r=\{0.5,1.5,\ldots,4.5$ fm\}, $\theta=\{0, \pi/8, \ldots, \pi/2\}$.

In a long history of structure study of 
$^8\textrm{Be}$ and  $^{12}\textrm{C}$, 
the $2\alpha$ and $3\alpha$ GCM calculations have been performed in many works
since 1970's (see Ref.~\cite{Fujiwara-supp,horiuchi-rev} and references therein), 
and successfully described cluster structures except for the ground state of  $^{12}\textrm{C}$. 
For the ground state of $^{12}$C,  the traditional $3\alpha$ models are not sufficient 
because the cluster breaking component, in particular, the $p_{3/2}$-closed configuration is 
significantly mixed in it, and therefore, they usually fail to reproduce 
the $0^+_1$-$2^+_1$ energy spacing and $B(E2;2^+_1\to 0^+_1)$. 
Suhara and the author have proposed an extended $3\alpha$ cluster model 
by adding the $p_{3/2}$-closed configuration in the $3\alpha$ GCM calculation, which we call 
the $3\alpha+p_{3/2}$ model \cite{Suhara:2014wua}. 
In the present calculation of  $^{12}\textrm{C}$, 
we apply the  $3\alpha+p_{3/2}$ model and take into account the cluster breaking component. 
We also apply the extended version, the $2\alpha+t+p_{3/2}$ model to  
 $^{11}\textrm{B}$ by taking account the cluster breaking component by 
adding the $(p_{3/2})^3_\pi (p_{3/2})^4$ configuration 
in the $2\alpha+t$ GCM calculation
(the $2\alpha+h+p_{3/2}$ model to  
 $^{11}\textrm{C}$ with  $(p_{3/2})^4_\pi (p_{3/2})^3$ configuration 
in the $2\alpha+h$ GCM calculation).

The nuclear density $\rho_N(r)$ in the core nuclei
is calculated for the obtained GCM wave function 
$\Psi(J^\pi_n)$. The $\rho_N(r)$ is the $r$-dependent spherical density of the $J^\pi_n$ state
after extraction of the cm motion. 

\subsection{Hamiltonian of nuclear part}
Hamiltonian of the nuclear part consists of the kinetic term, effective nuclear interactions, and Coulomb interactions
as follows, 
\begin{eqnarray}
H_N&=&T+V^\textrm{(c)}_N+V^\textrm{(so)}_N+V_\textrm{coul},\\
T&=&\sum^A_{i} \frac{1}{2m_N}\bvec{p}^2_i -T_G,\\
V^\textrm{(c)}_N&=&\sum^A_{i<j} v^{(c)}_{NN}(i,j),\\
V^\textrm{(so)}_N&=&\sum^A_{i<j} v^{\textrm{(so)}}_{NN}(i,j),\\ 
V_\textrm{coul}&=&\sum^Z_{i<j} v_\textrm{coul}(r_{ij}),
\end{eqnarray}
where  $T_G$ is the kinetic term of the cm motion, 
and  $v^{(c)}_{NN}(i,j)$ and  $v^{\textrm{(so)}}_{NN}(i,j)$ are the effective $N$-$N$ central and spin-orbit  interactions.
The energy $E_N$ of the core nucleus is given as
$E_N=\langle \Psi(J^\pi_n) |H_N|\Psi(J^\pi_n) \rangle$  (the nuclear energy). In the GCM calculation, 
the coefficients $c^{J^\pi_n}_{\bvec{S}_1,\ldots, \bvec{S}_k,K}$ in \eqref{eq:gcm-wf}  are determined so as to minimize $E_N$.

\subsection{Hamiltonian and folding potential of $\Lambda$-nucleus system}
$(0s)_\Lambda$ states of $\Lambda$ hypernuclei are calculated with a folding potential model
by solving the following single $S$-wave channel problem within local density approximations,
\begin{eqnarray}
H_{\Lambda}&=&T_{\Lambda}+U_\Lambda, \\
T_{\Lambda}&=&\frac{1}{2\mu_\Lambda}\bvec{p}^2, \\
\mu_\Lambda&=&\frac{(A-1)m_N m_\Lambda}{(A-1)m_N + m_\Lambda}, \\
U_\Lambda(\bvec{r},\bvec{r}')&=&U^\textrm{D}_\Lambda(\bvec{r})
+ | \bvec{r} \rangle  U^\textrm{EX}_\Lambda(\bvec{r},\bvec{r}')\langle \bvec{r}'|, \\
U^\textrm{D}_\Lambda(\bvec{r})&=&\int \bvec{r}'' \rho_N(\bvec{r}'') v^\textrm{D}_{\Lambda N}(k_f; |\bvec{r}-\bvec{r}''|),\\
U^\textrm{EX}_\Lambda(\bvec{r},\bvec{r}')&=&
\rho_N(\bvec{r},\bvec{r}') v^\textrm{EX}_{\Lambda N}(k_f; |\bvec{r}-\bvec{r}'|),\\
 v^\textrm{D}_{\Lambda N}(k_f; r)&=&\frac{1}{2}\left[ V^\textrm{e}_{\Lambda N}(k_f; r)+ V^\textrm{o}_{\Lambda N}(k_f; r)\right],\\
 v^\textrm{EX}_{\Lambda N}(k_f; r)&=&\frac{1}{2}\left[ V^\textrm{e}_{\Lambda N}(k_f; r)- V^\textrm{o}_{\Lambda N}(k_f; r)\right],
%\mu=\frac{(A-1)m_N m_\Lambda}{(A-1)m_N + m_\Lambda}
\end{eqnarray}
where $\bvec{r}$, $\bvec{r}'$, and $\bvec{p}$ are defined with respect to the relative coordinate of the $\Lambda$ 
from the cm of the core nucleus. 
$ V^\textrm{e}_{\Lambda N}(k_f; r)$  and  $V^\textrm{o}_{\Lambda N}(k_f; r)$ are the even and odd parts of the
effective $\Lambda$-$N$ central interactions, respectively, where $k_f$ is the parameter for density dependence of 
the effective $\Lambda$-$N$ interactions. 

The nuclear density matrix $\rho_N(\bvec{r},\bvec{r}')$ 
in the exchange potential  $U^\textrm{EX}_\Lambda(,\bvec{r}')$ is approximated
with the density matrix expansion (DME) using the LDA  \cite{Negele:1975zz},
\begin{eqnarray}
&&\rho_N(\bvec{r},\bvec{r}')\sim \rho^\textrm{DME}_N(\bvec{r},\bvec{r}'), \\
&&\rho^\textrm{DME}_N(\bvec{r},\bvec{r}')=\nonumber \\
&&\rho^\textrm{LDA}_N (\bvec{r},\bvec{r}')
\left(\frac{3}{k^\textrm{LDA}_f|\bvec{r}-\bvec{r}'|}  \right) j_1 (k^\textrm{LDA}_f|\bvec{r}-\bvec{r}'|),\\
&&\rho^\textrm{LDA}_N (\bvec{r},\bvec{r}')= \rho_N\left(\frac{\bvec{r}+\bvec{r}'}{2}\right),\\
&&k^\textrm{LDA}_f=\left[\frac {3\pi^2}{2}\rho^\textrm{LDA}_N (\bvec{r},\bvec{r}')  \right]^{1/3}.
\end{eqnarray}
To see ambiguity of choice of local density and Fermi momentum in the DME approximation
we also used the second choice (LDA2), 
\begin{eqnarray}
\rho^\textrm{LDA2}_N (\bvec{r},\bvec{r}')&=& \frac{1}{2}\left[\rho_N (\bvec{r})+\rho_N (\bvec{r}')\right ],\\
k^\textrm{LDA2}_f&=&\left[\frac {3\pi^2}{2}\rho^\textrm{LDA2}_N (\bvec{r},\bvec{r}')  \right]^{1/3},
\end{eqnarray}
and found that the first and the second choices give  qualitatively similar results. 
In this paper,  we use the DME approximation with the first choice in the calculation of the exchange folding 
potential $U^\textrm{EX}_\Lambda(\bvec{r},\bvec{r}')$.
%the density-independent and density-dependent effective $\Lambda$-$N$ interactions. 

For a given nuclear density $\rho_N(r)$, the $\Lambda$-core wave function $\phi_\Lambda(r)$ and 
energy $E_\Lambda= \langle\phi_\Lambda| H_\Lambda |\phi_\Lambda\rangle$ are calculated by solving
the one-body potential problem with the Gaussian expansion method \cite{Kamimura:1988zz,Hiyama:2003cu}. 
The rms radius ($r_\Lambda$) measured from the core nucleus and the averaged nuclear density ($\langle {\rho}_N\rangle_\Lambda$)
for the $\Lambda$ distribution are calculated with the obtained $\Lambda$-core wave function $\phi_\Lambda(r)$ , 
\begin{eqnarray}
r_\Lambda&=&\sqrt{\int  \phi_\Lambda^*(r)\phi_\Lambda(r) r^2 d\bvec{r}},\\
\langle {\rho}_N\rangle_\Lambda&=&\int  \phi_\Lambda^*(r)\phi_\Lambda(r) \rho_N(r)  d\bvec{r}.
\end{eqnarray}

\subsection{core polarization effect}
We take into account the core polarization, which is the structure change of core nuclei caused by 
the impurity, the $\Lambda$ particle,  in $\Lambda$ hypernuclei as follows.
In the present folding potential model, the $\Lambda$ binding reflects the core nuclear structure only through the 
nuclear density $\rho_N(r)$. When the $0s$-orbit $\Lambda$ particle is regarded as an impurity of the nuclear system, 
the $\Lambda$-$N$ interactions may act as an additional attraction to the nuclear system and make the nuclear size
slightly small.
To simulate the nuclear structure change induced by the $0s$-orbit $\Lambda$, we add artificial nuclear interactions 
by slightly enhancing the central part by hand 
and perform the GCM calculation of the nuclear system for the modified Hamiltonian,
\begin{equation}
H_N+\Delta H(\epsilon) =T+(1+\epsilon)V^\textrm{(c)}_N+V^\textrm{(so)}_N+V_\textrm{coul},\\
\end{equation}
with the additional term $\Delta H(\epsilon)=\epsilon V^\textrm{(c)}_N$, 
where $\epsilon$ is the enhancement factor and taken to be $\epsilon\ge 0$.
For the GCM wave function $\Phi(\epsilon;J^\pi_n)$ of the $J^\pi_n$ state obtained with 
$H_N+\Delta H(\epsilon) $, we calculate the nuclear energy $E_N(\epsilon)=\langle \Phi(\epsilon;J^\pi_n)| H_N |\Phi(\epsilon;J^\pi_n)\rangle$ and 
the nuclear density $\rho_N(\epsilon; r)$.
Then we  calculate 
 the $\Lambda$ wave function ($\phi_\Lambda(\epsilon;r)$) and 
energy ($E_\Lambda(\epsilon)$)  for the obtained  $\epsilon$-dependent nuclear density $\rho_N(\epsilon; r)$.
Finally, we search for the optimum  $\epsilon$  value 
so as to minimize the energy of the total system,
\begin{eqnarray}
E(\epsilon)&=&E_N(\epsilon)+E_\Lambda(\epsilon),\\
\frac{\delta E(\epsilon)}{\delta \epsilon}&=&0.
\end{eqnarray}
The $\Lambda$ binding energy ($B_\Lambda$) is calculated as 
$B_\Lambda=-(E(\epsilon)-E^\textrm{up}_N)$ for the optimized $\epsilon$ value, where 
$E^\textrm{up}_N=E_N(\epsilon=0)$ is the unperturbative nuclear energy without the $\Lambda$ particle.  

We vary only the GCM coefficients for the fixed basis cluster wave functions corresponding to the inert
cluster ansatz. In this assumption, the enhancement of the effective central nuclear interactions acts like an enhancement of the 
inter-cluster potentials.  

\section{Effective interactions}\label{sec:interactions}

\subsection{Effective nuclear interactions}

As for the effective two-body  nuclear interactions, we use the finite-range central interactions of
the Volkov No.2 parametrization \cite{VOLKOV} and the spin-orbit interactions of  the G3RS parametrization \cite{LS},
\begin{eqnarray}
v^\textrm{(c)}_{NN}(1,2)&=&V^\textrm{(c)}_{NN}(r_{12})(w+bP_\sigma-hP_\tau-mP_\sigma P_\tau), \\
V^\textrm{(c)}_{NN}(r)&=&v_1\exp\left[-\left(\frac{r}{a_1}\right)^2\right]+v_2\exp\left[-\left(\frac{r}{a_2}\right)^2\right],
\nonumber\\ \\
v_1&=&-60.65\ \textrm{MeV}, \ v_2=61.14\ \textrm{MeV},\\
a_1&=&1.80\   \textrm{fm},\  a_2=1.01 \  \textrm{fm},\\
v^\textrm{(so)}_{NN}(1,2)&=&V^\textrm{(so)}_{NN}(r) \frac{1+P_\sigma}{2}\frac{1+P_\sigma P_\tau}{2} (\bvec{l}_{12}\cdot \bvec{s}_{12}),\\
V^\textrm{(so)}_{NN}(r)&=&u_1\exp\left[-\left(\frac{r}{b_1}\right)^2\right]
+u_2\exp\left[-\left(\frac{r}{b_2}\right)^2\right],
\nonumber\\ \\
b_1&=&0.60\   \textrm{fm},\ b_2=0.447\   \textrm{fm},
\end{eqnarray}
where $P_\sigma$($P_\tau$) is the spin(isospin) exchange operator, $r_{12}$ is the relative distance
$r_{12}=|\bvec{r}_{12}|$ for the relative coordinate $\bvec{r}_{12}=\bvec{r}_1-\bvec{r}_2$,  
$\bvec{l}_{12}$ is the angular momentum for $\bvec{r}_{12}$, and $\bvec{s}_{12}$ is the sum of nucleon spins 
$\bvec{s}_{12}=\bvec{s}_1+\bvec{s}_2$. 

We use $w=0.40$, $m=0.60$, and $b=h=0.125$ for the central interactions, and $u_1=-u_2=1600$ MeV
for the spin-orbit interactions.  These parameters reproduce
the deuteron binding energy, the $\alpha$-$\alpha$ scattering phase shift, and properties of the ground and excited states of 
$^{12}$C \cite{Suhara:2014wua,uegaki1,uegaki3}.
For $^6$Li, we use a modified values 
$w=0.43$, $m=0.57$, $b=h=0.125$, and $u_1=-u_2=1200$ MeV 
to reproduce the $^6\textrm{Li}(1^+_1)$ and  $^6\textrm{Li}(3^+_1)$  energies
relative to the $\alpha+d$ threshold energy. Note that this modification gives no effect on 
$s$-shell nuclei, $d$, $t$, $h$, and $\alpha$.

\subsection{Effective $\Lambda$-nucleon interactions}
For the effective $\Lambda$-$N$ central interactions, 
we use the %density-dependent 
$G$-matrix interactions derived from $\Lambda$-$N$ interactions of the one-boson-exchange model, which we denote as the $\Lambda NG$ interactions
\cite{Yamamoto:2010zzn,Rijken:2010zzb}. In this paper, we adopt the central part of the $\Lambda NG$ interactions with the ESC08a parametrization,
\begin{eqnarray}
V^\textrm{e}_{\Lambda N}(k_f; r)&=&\sum^3_i  (c^\textrm{e}_{0,i}+c^\textrm{e}_{1,i}  k_F+c^\textrm{e}_{2,i}  k_F^2)  \exp\left[-\left(\frac{r}{\beta_i}\right)^2\right],
\nonumber\\
&&\\
V^\textrm{o}_{\Lambda N}(k_f; r)&=&\sum^3_i  (c^\textrm{o}_{0,i}+c^\textrm{o}_{1,i}  k_F+c^\textrm{o}_{2,i}  k_F^2)  \exp\left[-\left(\frac{r}{\beta_i}\right)^2\right],
\nonumber\\
&&\\
c^\textrm{e}_{n,i}&=&\frac{1}{4}c^\textrm{1E}_{n,i}+\frac{3}{4}c^\textrm{3E}_{n,i},\\
c^\textrm{o}_{n,i}&=&\frac{1}{4}c^\textrm{1O}_{n,i}+\frac{3}{4}c^\textrm{3O}_{n,i},
\end{eqnarray}
with $\beta_1=0.5$ fm, $\beta_2=0.9$ fm,  and $\beta_3=2.0$ fm. Values of the parameters $c^\textrm{1E,3E,1O,3O}_{n,i}$ are listed in Table \ref{tab:ESC08a}.
Note that, in the present $S$-wave $\Lambda$ calculation, the effective $\Lambda$-$N$ interactions are spin-independent central interactions, as 
the singlet and triplet parts are averaged with the factors 1/4 and 3/4, respectively, 
and the spin-orbit interactions are dropped off. 
%Consequently, the spin partner states  completely degenerate.   

As for the $k_f$ parameter of the $\Lambda NG$ interactions,  
we adopt two treatments. 
One is the density-dependent  interactions with $k_f=\langle k_f \rangle_\Lambda$, where 
$\langle k_f \rangle_\Lambda$ is the averaged Fermi momentum for the $\Lambda$ particle,
\begin{equation}
\langle k_f \rangle_\Lambda
=\left[\frac {3\pi^2}{2}\langle \rho_N \rangle_\Lambda \right]^{1/3},
\end{equation}
and self-consistently determined for each state. This $k_f$ choice of the $\Lambda NG$ interactions 
is the so-called ``averaged density approximation (ADA)''  
used in Refs.~\cite{Yamamoto:2010zzn,Isaka:2016apm,Isaka:2017nuc}.
The other is the density-independent interaction with a 
fixed $k_f$ value, $k_f=k^\textrm{inp}_f$. Here, the input parameter $k^\textrm{inp}_f$
is chosen for each $^A_\Lambda Z$ system. It means that the  $k^\textrm{inp}_f$  is system dependent but 
``state independent''. In this paper, we use the mean value of 
$\langle k_f \rangle_\Lambda$ of low-energy states obtained by the former treatment (ADA) as the input of $k^\textrm{inp}_f$. 
These choices reasonably reproduce the  $\Lambda$ binding energies of 
$^{7}_\Lambda\textrm{Li}$, $^{9}_\Lambda\textrm{Be}$, $^{11}_\Lambda\textrm{B}$,
$^{12}_\Lambda\textrm{B}$, $^{12}_\Lambda\textrm{C}$, and $^{13}_\Lambda\textrm{C}$. 
We label the first treatment, density-dependent $\Lambda NG$ interactions  with $k_f=\langle k_f \rangle_\Lambda$, 
as ESC08a(DD),  and the second one, the density-independent  $\Lambda NG$  interactions with 
$k_f=k^\textrm{inp}_f$  ESC08a(DI). Note that the former is state-dependent (structure-dependent) and the latter is 
state-independent (structure-independent), but 
the system-dependent $k_f$ is used in both cases.

The $\Lambda NG$ interactions have been applied to various structure model calculations of hypernuclei 
such as cluster model, mean-field, and HAMD calculations. 
In the application of the $\Lambda NG$ interactions to cluster model and HAMD calculations, 
the parameter $k_f$ of the  density-independent $\Lambda NG$ interactions  is usually 
adjusted to fit the $\Lambda$ binding energy for each (sub)system.
In applications of the effective $\Lambda NG$ interactions to $^A_\Lambda Z$ in a wide mas number region, 
the density-dependent $\Lambda NG$ interactions have been used, for instance, 
in the mean-field calculations and recent HAMD calculations \cite{Isaka:2016apm,Isaka:2017nuc}, 
because they were originally 
designed in the density-dependent form to reproduce systematics of $\Lambda$ binding energy \cite{Yamamoto:2010zzn}.
%The density-dependent $\Lambda NG$ interactions have been used
%in mean-field calculations \cite{Tamamoto2010} and also in the recent Hyper-AMD calculations \cite{Isaka:2016apm,Isaka:2017nuc}. 
In Refs.~\cite{Yamamoto:2010zzn,Isaka:2016apm}, they also 
showed the results with another choice of $k_f=k^\textrm{LDA2}_f$ in addition to the ADA results. 
In the present calculation for $(0s)_\Lambda$ states, the results obtained with
$k_f=k^\textrm{LDA2}_f$ show similar results to the present 
ESC08a(DD) ones.
%show trends similar to those with the LDA.

\begin{table}[ht]
\caption{Parameters of the $\Lambda NG$ interactions of ESC08a from Table II of Ref.~\cite{Yamamoto:2010zzn}.
\label{tab:ESC08a} }
\begin{center}
\begin{tabular}{ccccccc}
\hline
& $i=1$ & $i=2$ & $i=3$ \\
$c^\textrm{1E}_{0,i}$	& $	-3144	$ & $	368.0 	$ & $	-1.467	$ \\
$c^\textrm{1E}_{1,i}$	& $	6411	$ & $	-984.4	$ & $	0	$ \\
$c^\textrm{1E}_{2,i}$	& $	-2478	$ & $	394.5	$ & $	0	$ \\
	& $		$ & $		$ & $		$ \\
$c^\textrm{3E}_{0,i}$	& $	-2734	$ & $	316.8	$ & $	-1.044	$ \\
$c^\textrm{3E}_{1,i}$	& $	5827	$ & $	-901.6	$ & $	0	$ \\
$c^\textrm{3E}_{2,i}$	& $	-2404	$ & $	395.8	$ & $	0	$ \\
	& $		$ & $		$ & $		$ \\
$c^\textrm{1O}_{0,i}$	& $	663.1	$ & $	124.6	$ & $	-0.5606	$ \\
$c^\textrm{1O}_{1,i}$	& $	1728	$ & $	-50.97	$ & $	0	$ \\
$c^\textrm{1O}_{2,i}$	& $	-599	$ & $	32.4	$ & $	0	$ \\
	& $		$ & $		$ & $		$ \\
$c^\textrm{3O}_{0,i}$	& $	810.6	$ & $	-182.7	$ & $	-0.7257	$ \\
$c^\textrm{3O}_{1,i}$	& $	-703.2	$ & $	118.1	$ & $	0	$ \\
$c^\textrm{3O}_{2,i}$	& $	209.6	$ & $	-13.17	$ & $	0	$ \\
\hline		
\end{tabular}
\end{center}
\end{table}

\section{Results}\label{sec:results}
By applying the $\alpha+d$, $2\alpha$, $2\alpha+d$, $2\alpha+t(h)+p_{3/2}$, and  $3\alpha+p_{3/2}$ 
GCM to core nuclei, $^6\textrm{Li}$, $^8\textrm{Be}$, $^{10}\textrm{B}$, $^{11}\textrm{B(C)}$,
and $^{12}\textrm{C}$, we calculate 
$(0s)_\Lambda$ states in $^A_\Lambda Z$ with the
single-channel folding potential model by taking account the core polarization effect.
% using the ESC08a(DI) and ESC08a(DD) interactions.

In the present calculation,  the $\Lambda$ particle around the 
$I^\pi$ state of the core nucleus $^{A-1} Z(I^\pi)$ feels
the spin-independent potentials, and therefore the spin partner $J^\pi= (I\pm 1/2)^\pi$ states in $^A_\Lambda Z$
completely degenerate.
We denote the spin partner $J^\pi= (I\pm 1/2)^\pi$ states  
in $\Lambda$ hypernuclei by $^{A}_\Lambda{Z}(I^\pi)$.
We calculate low-lying $I^\pi$ states with dominant $0\hbar\omega$ configurations in $^{A-1}{Z}$, and the corresponding 
$(0s)_\Lambda$ states in $^{A}_\Lambda{Z}$.  

The $^8\textrm{Be}(0^+_1)$ and $^6\textrm{Li}(3^+_1)$ states, which are strictly speaking 
quasi-bound states, are calculated in the bound state approximation
with the boundary condition $d\le 15$ fm of the GCM model space. 
The present GCM calculation gives stable results
for these states. The $^8\textrm{Be}(2^+_1)$ state is a broad resonance state, for which we can not
obtain a stable result in the bound state approximation.  Instead, we calculate the excitation energy of 
$^8\textrm{Be}(2^+_1)$ from the $\alpha+\alpha$ scattering  phase shifts
 with the resonating group method (RGM).

To see the effect of the cluster breaking component in $^{A-1}Z$ and $^A_\Lambda Z$,
we also show some results for $^{12}\textrm{C}$ and $^{13}_\Lambda\textrm{C}$
obtained by the traditional $3\alpha$ GCM calculation 
without the cluster breaking ($p_{3/2}$) component and compare them 
with those obtained by the present 
$3\alpha+p_{3/2}$ model.
Note that, in the present model,  the cluster breaking components
contribute only to $^{12}\textrm{C}(0^+)$
and  $^{11}\textrm{B,C}(3/2^-)$ but do not affect other spin-parity states.

\subsection{Properties of core nuclei}

\begin{table}[ht]
\caption{Energies  (MeV), radii (fm), and $B(E2)$ (e$^2$fm$^{4}$)  in ordinary nuclei. 
Binding energies $(-E_N)$, relative energies $(E_r)$ measured from the cluster-decay threshold, 
calculated rms matter radii ($R_N$), the experimental rms point-proton radii $(R_p)$,
and $E2$ transition strengths to the ground states are listed. 
%The calculated values for $s$-shell nuclei are those obtained for $(0s)$ configurations with the 
%fixed width parameter  $\nu=0.235$ fm$^{-2}$. For $^{12}\textrm{C}$, the values calculated with the 
%traditional 3$\alpha$-cluster model without the $(3/2)^8$ configuration are shown for comparison. 
For $^{12}\textrm{C}$, the results obtained with the present $3\alpha+p_{3/2}$ model 
and those with the $3\alpha$ model without the $p_{3/2}$ component are shown.
The experimental data are from Refs.~\cite{AjzenbergSelove:1990zh,Tilley:2002vg,Tilley:2004zz,Kelley:2012qua}.
\label{tab:nuclei} }
\begin{center}
\begin{tabular}{ccccccc}
\hline
 & \multicolumn{2}{c}{$-E_N$} &  \multicolumn{2}{c}{$E_r$} & $R_N$ & $R_p$ \\
	&	cal	&	exp	&	cal	&	exp	&	cal	&	exp	\\
$d(1^+_1)$	&	0.43 	&	2.224	&		&		&	0.98 	&	1.941 	\\
$t(1/2^+_1)$	&	6.9 	&	8.481	&		&		&	1.23 	&	1.504 	\\
$\alpha(0^+_1)$	&	27.6 	&	28.296	&		&		&	1.55 	&	1.410 	\\
$^6\textrm{Li}(1^+_1)$	&	29.5 	&	31.995	&	$-$1.48 	&	$-$1.48 	&	2.56 	&	2.426 	\\
$^8\textrm{Be}(0^+_1)$	&	55.0 	&	56.499	&	0.21 	&	0.09 	&	3.37 	&		\\
$^{10}\textrm{B}(3^+_1)$	&	60.5 	&	64.75	&	$-$4.88 	&	$-$5.93 	&	2.39 	&	2.253 	\\
$^{11}\textrm{B}(3/2^-_1)$	&	71.8 	&	76.203	&	$-$9.66 	&	$-$11.13 	&	2.33 	&	2.229 	\\
$^{11}\textrm{C}	(3/2^-_1)$&	69.2 	&	73.439 	&	$-$7.05 	&	$-$8.37 	&	2.34 	&		\\
$^{12}\textrm{C}(0^+_1)$	&	90.2	&	92.16	&	$-$7.37&	$-$7.27 	&	2.35 	&	2.298 	\\
w/o $p_{3/2}$ &	88.1	&	&	$-$3.22&&	2.52 	&		\\
& \multicolumn{2}{c}{$B(E2)$}& 		&		&		&		\\
	&	cal	&	exp	&		&	&	&\\
$^6\textrm{Li}(3^+_1)$	&	11.3 	&	10.7(8)	&		&		&		&		\\
$^{10}\textrm{B}(1^+_1)$	&	5.2 	&	4.15(2)	&		&		&		&		\\
$^{11}\textrm{B}(5/2^-_1)$	&	9.5 	&	8.9(3.2)	&		&		&		&		\\
$^{12}\textrm{C}(2^+_1)$&	7.3 	&	7.6(4)	&		&		&		&		\\
w/o $p_{3/2}$ &	10.6	&7.6(4)		&		&&	&		\\
\hline		
\end{tabular}
\end{center}
\end{table}

\begin{table}[ht]
\caption{The calculated values of rms radii ($R_N$ (fm)), the size difference ($R_N-R_{N,\textrm{gs}}$ (fm)), excitation energies
in $^{11}\textrm{B}$, and energy difference from the mirror nucleus $^{11}\textrm{C}$. The difference 
in the binding energy ($\Delta_\textrm{mir}(-E_N)$ (MeV)) for the ground state and that in the excitation energies 
 ($\Delta_\textrm{mir}(E_x)$ (MeV)) for excited states are shown.  The experimental data are taken from Ref.~\cite{Kelley:2012qua}.
\label{tab:shift} }
\begin{center}
\begin{tabular}{ccccccc}
\hline
	&  $R_N$ & &  & 
& \multicolumn{2}{c}{$\Delta_\textrm{mir}$(B.E) }\\
 &	&  	&	& &cal	&	exp	\\
$^{11}\textrm{B}(3/2^-_1)$	&	2.33 	&	&		&	&	2.60 	&	2.764 	\\
	&  $R_N$ &	$R_N$-$R_{N,\textrm{gs}}$& \multicolumn{2}{c}{$E_x$ }  & 
 \multicolumn{2}{c} {$\Delta_\textrm{mir}(E_x)$ } \\
	&		&				&	cal & exp	&	cal	&	exp	\\
$^{11}\textrm{B}(1/2^-_1)$	&	2.50 	&	0.18 	&	2.79 	&	2.125 	&	0.17 	&	0.13 	\\
$^{11}\textrm{B}(3/2^-_2)$	&	2.58 	&	0.26 	&	5.57 	&	5.020 	&	0.22 	&	0.22 	\\
$^{11}\textrm{B}(5/2^-_1)$	&	2.51 	&	0.19 	&	4.66 	&	4.445 	&	0.16 	&	0.13 	\\
\hline		
\end{tabular}
\end{center}
\end{table}

Nuclear properties of isolate core nuclei without the $\Lambda$ particle are shown in Tables \ref{tab:nuclei} and \ref{tab:shift}.  
The calculated values of the binding energies $(-E_N)$,  relative energies ($E_r$)  measured from cluster break-up threshold energies, root-mean-square (rms)  radii of nuclear matter ($R_N$), and $E2$ transition 
strengths  to the ground states are listed compared with experimental data in Table \ref{tab:nuclei}. 
For the experimental data of nuclear radii, the rms radii of point-proton distribution  ($R_p$) reduced from the charge radii are shown.
We also show the results for
$d$, $t$, and $\alpha$ clusters of $(0s)$ configurations with $\nu=0.235$ fm$^{-2}$. 
The energies and sizes are reasonably reproduced by the calculation except for the deuteron and triton.
The deuteron size is much underestimated, 
because the fixed-width $(0s)^2$ configuration is assumed in the present cluster model.
The calculated $B(E2)$  are in agreement with the experimental data without using any effective charges. 
For $^{12}$C, the $3\alpha$ GCM calculation without the cluster breaking ($p_{3/2}$) gives a larger size and 
$B(E2;2^+_1\to 0^+_1)$ than those of the present calculation,  
meaning that $^{12}\textrm{C}(0^+_1)$ slightly shrinks because of the cluster breaking effect
as discussed in Ref.~\cite{Suhara:2014wua}.

In Table  \ref{tab:shift}, we show the  Coulomb shift $\Delta_\textrm{mir}(E_x)$, which is 
defined by the excitation energy difference between mirror nuclei, 
for $A=11$ nuclei  together with calculated radii. 
Since the Coulomb interactions 
give only minor change of nuclear structure, and therefore the Coulomb shift sensitively probes 
the size difference between the ground and excited states except for weakly bound or resonance states. 
The calculated Coulomb shifts for $^{11}\textrm{B}(1/2^-_1, 3/2^-_2, 5/2^-_1)$  agree well with the experimental data indicating that
the size differences of these states are reasonably described by the present calculation. 

\subsection{Ground states of $\Lambda$ hypernuclei}
We here discuss the ground state properties of $^A_\Lambda Z$. 
In $^{A}_\Lambda{Z}$, the nuclear size $R_N$ ($R_N$ is the rms nuclear matter radius measured from the cm of the core nucleus) slightly decreases and the nuclear energy $E_N$ slightly increases from 
the original size ($R^\textrm{up}_N$) 
and energy ($E^\textrm{up}_N$) of unperturbative core nuclei $^{A-1}{Z}$ without the $\Lambda$.
We calculate the nuclear size change $\delta_\Lambda(R_N)=R_N-R^\textrm{up}_N$ 
and the nuclear energy change $\delta_\Lambda(E_N)=E_N-E^\textrm{up}_N$ caused by the $\Lambda$ particle in $^{A}_\Lambda{Z}$. 
To see the core polarization effect, we also calculate the $\Lambda$ energy gain,
$\Delta_\textrm{cp}(E_\Lambda)=E_\Lambda(\epsilon)-E_\Lambda(\epsilon=0)$, defined by 
the energy difference between the calculations 
with and without the core polarization. Here $E_\Lambda(\epsilon=0)$ is the $\Lambda$ energy without the core polarization,
that is the $\Lambda$ energy in the $\Lambda$-($^{A-1}Z)$ system with the unperturbative core nucleus. 

In Table \ref{tab:L-nuclei-gs}, we show  the calculated results of the ground state properties of $^{A}_\Lambda{Z}$ together with the experimental 
$B_\Lambda$.  As reference data, we also show the results for $^5_\Lambda\textrm{He}$ 
obtained by the $\Lambda$-$\alpha$ calculation with the inert $\alpha$ core assumption. 
Systematics of $\Lambda$ binding energies in this mass-number region is reasonably reproduced in both ESC08a(DI) and ESC08a(DD) interactions,  
though the reproduction is not perfect.

For $A>10$ systems, the nuclear size change $\delta_\Lambda(R_N)$ is less than 5\%. 
The small size change of the core nucleus in the ground state of $^{13}_\Lambda\textrm{C}$ is consistent with
the prediction of other calculations \cite{Motoba:1984ri, motoba85,Hiyama:1997ub,Hiyama:2010zzc,Funaki:2017asz}. 
Moreover, the nuclear energy change 
$\delta_\Lambda(E_N)$ and $\Lambda$ energy gain $\Delta_\textrm{cp}(E_\Lambda)$ by the core polarization 
are also small and compensate each other. 
It  indicates that
the core polarization effect  is minor and regarded as a higher-order perturbation in the $\Lambda$ binding except for $A<10$ systems.
The core polarization effects in the ESC08a(DD) results for $A>10$ systems are particularly small, because 
the ESC08a(DD) interactions become weak as the nuclear density increases because of the $k_f$ dependence. 
%Namely, the $k_f$ dependence of 
%the ESC08a(DD) interactions suppresses the size reduction of the core nuclei.

\begin{figure}[!h]
\begin{center}
\includegraphics[width=5.0cm]{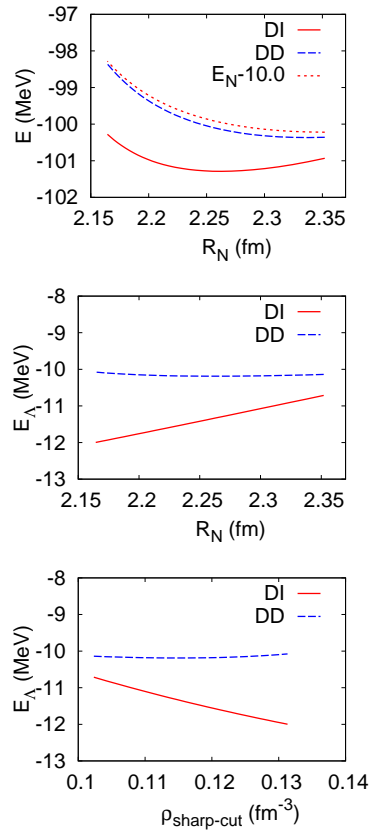} 	
\end{center}
%\vspace{0.5cm}
  \caption{(color online) 
Total energy ($E(\epsilon)=E_N(\epsilon)+E_\Lambda(\epsilon)$) and 
 $\Lambda$ energy $(E_\Lambda(\epsilon))$ for 
polarized core $\Phi(\epsilon)$ in  $^{13}_\Lambda\textrm{C}(0^+_1)$. The energies are 
plotted against the rms nuclear matter radius $R_N(\epsilon)$ in the top and middle panels.
The nuclear energy ($E_N(\epsilon)$) subtracted by $10$ MeV is also shown.
$E_\Lambda(\epsilon)$ plotted to the sharp-cut density $\rho_\textrm{sharp-cut}=(3/4\pi)(3/5)^{3/2}A R_N^{-3}$ 
reduced from $R_N(\epsilon)$ for the uniform density ansatz is shown in the bottom panel. 
The calculated values obtained with ESC08a(DI) and ESC08a(DD) are shown. 
%Energies are measured from the ground state energy obtained by the full sAMD+GCM.
\label{fig:r-dep}}
\end{figure}

As explained previously, the core polarization effect is taken into account 
by changing the enhancement factor $\epsilon$, which  
can be regarded as a control parameter of the nuclear size $R_N$.
In Fig.~\ref{fig:r-dep}, we show the nuclear size dependence of 
$E_N(\epsilon)$, $E_\Lambda(\epsilon)$, and $E(\epsilon)=E_N(\epsilon)+E_\Lambda(\epsilon)$ in $^{13}_\Lambda\textrm{C}$
obtained by varying the enhancement factor $\epsilon$.
The energies are plotted as functions of  the nuclear size $R_N(\epsilon)$. The $R_N^{-3}$-dependence of $E_\Lambda(\epsilon)$ is also shown.
In the ESC08a(DI) result, the $\Lambda$ energy ($E_\Lambda$) gradually goes down with the nuclear size reduction because the higher nuclear density 
gives larger attraction to the  $\Lambda$ potentials. 
As a result, the $\Lambda$ particle 
slightly reduces the core nuclear size. In contrast, in the ESC08a(DD) result, the $\Lambda$ energy has almost no dependence on the nuclear size, 
because the density-dependence of the $\Lambda NG$ interactions compensates the $\Lambda$ energy gain in the higher nuclear density.
As a result, the $\Lambda$ particle hardly changes the core nucleus size. 
Namely, the density dependence of the ESC08a(DD) interactions suppresses the size reduction of the core nuclei. 

Let us turn to $A<10$ systems, $^7_\Lambda\textrm{Li}$ and $^9_\Lambda\textrm{Be}$. Differently from $A>10$ systems, 
rather significant core size reduction occurs, because $^6\textrm{Li}$ and $^8\textrm{Be}$ have spatially 
developed $\alpha+d$- and $2\alpha$-cluster structures, respectively, and they are 
rather fragile (soft) against the size reduction. This is consistent with the size shrinkage predicted by 
pioneering works in Refs.~\cite{Motoba:1984ri,motoba85} followed by 
many works (see a review paper \cite{Hiyama:2010zzc} and references therein). Particularly remarkable core polarization effects are found in $^9_\Lambda\textrm{Be}$, 
because $^8\textrm{Be}$ is a very fragile system of the loosely bound (strictly speaking, quasi-bound) 
$2\alpha$ state.
The core polarization effects are  seen in the nuclear size change $\delta_\Lambda(R_N)$ and also 
the energy changes, $\delta_\Lambda(E_N)$ and $\delta_\Lambda(E_\Lambda)$, in both ESC08a(DI) and ESC08a(DD) calculations. 
For $^7_\Lambda\textrm{Li}$, the core size reduction is 13\% in the ESC08a(DI)  result, whereas it is 6\% in the ESC08a(DD) result.
It should be commented that the size reduction discussed here is the reduction of  nuclear matter radii of  core nuclei.
%On the other hand, the so-called size shrinkage in  $^7_\Lambda\textrm{Li}$ and $^9_\Lambda\textrm{Be}$ has been often discussed for the reduction of 
%the inter-cluster distance.
% The size shrinkage for the $\alpha$-$d$ distance in the present ESC08a(DI) result is 14\%, which is 
%almost consistent with the theoretical values predicted in Refs.\cite{Motoba:1984ri,motoba85,hiyama1999}. 
Detailed discussions of the shrinkage of the inter-cluster distance in  $^7_\Lambda\textrm{Li}$ and $^9_\Lambda\textrm{Be}$ are given later. 

%As shown later, the significant size change occurs also in the excited state $^7_\Lambda\textrm{Li}(3^+)$ and the $B(E2;3^+\to 1^+)$ of the core nucleus in 
%$^7_\Lambda\textrm{Li}
% The 3-body and 4-body cluster-model calculations of  $^7_\Lambda\textrm{Li}$ \cite{} have predicted about 20\% size reduction, which has been supported by
%the $\gamma$-ray spectroscopy experiments \cite{}. 

\begin{table*}[ht]
\caption{Ground state properties of $\Lambda$ hypernuclei. 
The $\Lambda$ distribution size ($r_\Lambda$ (fm)), averaged Fermi momentum  ($\langle k_f \rangle_\Lambda$ fm$^{-1}$), core nuclear size ($R_N$ (fm)), 
nuclear size change ($\delta_\Lambda(R_N)$ (fm)),
nuclear energy change ($\delta_\Lambda(E_N)$ (MeV)), difference  of $\Lambda$ energy with and without core polarization ($\Delta_\textrm{cp}(E_\Lambda)$ (MeV)), and 
the $\Lambda$ binding energy ($B_\Lambda$ (MeV)) are listed. 
The calculated results obtained with
ESC08a(DI) and ESC08a(DD) are shown. 
The experimental $B_\Lambda$ values are taken from the data compilation in Ref.~\cite{Davis:2005mb}.
The experimental data of spin-averaged values 
($\overline{B}_\Lambda$ (MeV))) 
of the $\Lambda$ binding energy for spin partners, $J^\pi=I^\pi \pm 1/2$, are also 
shown. 
\label{tab:L-nuclei-gs} }
\begin{center}
\begin{tabular}{ccccccccccc}
\hline
\multicolumn{11}{c}{ESC08a(DI)}	\\
	&$k^\textrm{inp}_f$	& $r_\Lambda$ &	$\langle k_f \rangle_\Lambda$	&	$R_N$	&		$\delta_\Lambda({R_N})$	&	$\delta_\Lambda({E_N})$	&
$\Delta_\textrm{cp}(E_\Lambda)$&	$B_\Lambda$	&	$B_{\Lambda,\textrm{exp}}$	 & $\overline{B}_{\Lambda,\textrm{exp}}$ \\
$^5\textrm{He}(0^+)$	&	0.95 	&	2.84 	&	0.95 	&	1.55 	&		&		&		&	3.6 	&	3.12(2)	&	3.12(2)	\\
$^7\textrm{Li}(1^+)$	&	0.93 	&	2.57 	&	0.95 	&	2.22 	&$	-0.33 	$&	0.59 	&$	-0.85 	$&	5.4 	&	5.58(3)	&	5.12(3)	\\
$^9\textrm{Be}(0^+)$	&	0.90 	&	2.44 	&	0.98 	&	2.44 	&$	-0.94 	$&	1.69 	&$	-3.15 	$&	7.0 	&	6.71(4)	&	6.71(4)	\\
$^{11}\textrm{B}(3^+)$	&	1.03 	&	2.36 	&	1.10 	&	2.29 	&$	-0.10 	$&	0.37 	&$	-0.42 	$&	10.0 	&	10.24(5)	&	10.09(5)	\\
$^{12}\textrm{B}(3/2^-)$	&	1.07 	&	2.33 	&	1.16 	&	2.24 	&$	-0.09 	$&	0.29 	&$	-0.36 	$&	10.9 	&	11.37(6)	&	11.27(6)	\\
$^{12}\textrm{C} (3/2^-)$	&	1.06 	&	2.32 	&	1.16 	&	2.25 	&$	-0.09 	$&	0.31 	&$	-0.42 	$&	11.1 	&	10.76(19)	&	10.65(19)	\\
$^{13}\textrm{C}(0^+)$	&	1.11 	&	2.35 	&	1.18 	&	2.26 	&$	-0.09 	$&	0.27 	&$	-0.36 	$&	11.1 	&	11.69(12)	&	11.69(12)	\\
w/o $p_{3/2}$	&	1.11 	&	2.45 	&	1.11 	&	2.41 	&$	-0.11 	$&	0.35 	&$	-0.42 	$&	9.9 	&	11.69(12)	&	11.69(12)	\\
\multicolumn{11}{c}{ESC08a(DD)}	\\
	& 	& $r_\Lambda$ &	$\langle k_f \rangle_\Lambda$	&	$R_N$	&		$\delta_\Lambda({R_N})$	&	$\delta_\Lambda({E_N})$	&
$\Delta_\textrm{cp}(E_\Lambda)$&	$B_\Lambda$	&	$B_{\Lambda,\textrm{exp}}$	 & $\overline{B}_{\Lambda,\textrm{exp}}$ \\
$^5\textrm{He}(0^+)$	&		&	2.83 	&	0.95 	&	1.55 	&		&		&		&	3.6 	&	3.12(2)	&	3.12(2)	\\
$^7\textrm{Li}(1^+)$	&		&	2.66 	&	0.91 	&	2.40 	&$	-0.15 	$&	0.08 	&$	-0.12 	$&	5.4 	&	5.58(3)	&	5.12(3)	\\
$^9\textrm{Be}(0^+)$	&		&	2.67 	&	0.90 	&	2.69 	&$	-0.68 	$&	0.44 	&$	-1.19 	$&	6.4 	&	6.71(4)	&	6.71(4)	\\
$^{11}\textrm{B}(3^+)$	&		&	2.48 	&	1.06 	&	2.38 	&$	-0.01 	$&	0.00 	&$	0.00 	$&	9.0 	&	10.24(5)	&	10.09(5)	\\
$^{12}\textrm{B}(3/2^-)$	&		&	2.45 	&	1.11 	&	2.33 	&$	0.00 	$&	0.00 	&$	0.00 	$&	9.6 	&	11.37(6)	&	11.27(6)	\\
$^{12}\textrm{C} (3/2^-)$	&		&	2.45 	&	1.11 	&	2.34 	&$	0.00 	$&	0.00 	&$	0.00 	$&	9.6 	&	10.76(19)	&	10.65(19)	\\
$^{13}\textrm{C}(0^+)$	&		&	2.44 	&	1.13 	&	2.35 	&$	0.00 	$&	0.00 	&$	0.00 	$&	10.1 	&	11.69(12)	&	11.69(12)	\\
w/o $p_{3/2}$	&		&	2.47 	&	1.08 	&	2.51 	&$	-0.01 	$&	0.01 	&$	-0.01 	$&	10.1 	&	11.69(12)	&	11.69(12)	\\
\hline		
\end{tabular}
\end{center}
\end{table*}

%%%%%%%%%%%%%%%%%%%%%%%%%%%%%%
%%%%%%%%%%%%%%%%%%%%%%%%%%%%%

\subsection{Excited states of  $\Lambda$ hypernuclei}
We also apply the present method to core-excited $(0s)_\Lambda$ states in $\Lambda$ hypernuclei.
A particular attention is paid to excitation energy shift and its 
relation to nuclear size difference from
the ground state in each $^{A}_\Lambda{Z}$ system.

\begin{table*}[ht]
\caption{Properties of excited states in $\Lambda$ hypernuclei. 
The $\Lambda$ distribution size ($r_\Lambda$ (fm)),  averaged Fermi momentum  ($\langle k_f \rangle_\Lambda$ (fm$^{-1}$)), core nuclear size ($R_N$ (fm)), 
nuclear size change ($\delta_\Lambda(R_N)$ (fm)), the difference  $R_{N}-R_{N,\textrm{gs}}$ (fm) 
of the nuclear size from that of the ground state, excitation energies in $^{A-1}Z$ and  $^{A}_\Lambda Z$ systems ($^{A-1} E_x$ and $^A_\Lambda E_x$ (MeV)), 
and the excitation energy shift $\delta_\Lambda(E_x)$ (MeV). The calculated values obtained  with ESC08a(DI) are shown together with  
$\delta_\Lambda(E_x)$ calculated with ESC08a(DD).
For details of  the experimental data of excitation energies, see the caption of Fig.~\ref{fig:spe1}. 
%The experimental values of the excitation energy shift 
%$\delta_\Lambda({E_{x}})$ are reduced from experimental excitation energies in $^{A-1}Z$ from Refs.\cite{AjzenbergSelove:1990zh,Tilley:2002vg,Tilley:2004zz,Kelley:2012qua}
%and those in  $^A_\Lambda Z$ from Refs.~\cite{Tamura:2010zz,Tanida:2000zs,Ukai:2006zp,Akikawa:2002tm,Miura:2005mh,Ma:2010zzb,Ajimura:2001na,Kohri:2001nc,Tang:%2014atx,Hosomi:2015fma}. 
%In the evaluation of the experimental $\delta_\Lambda({E_{x}})$ values in   $^A_\Lambda Z$,  
 %the spin-averaged excitation energies are used except for 
%$^{12}_\Lambda\textrm{B}(1/2^-_1,3/2^-_2)$,and $^{12}_\Lambda\textrm{C}(1/2^-_1,3/2^-_2)$,
%and   $^{13}_\Lambda\textrm{C}(2^+)$. For $^{12}_\Lambda\textrm{B(C)}(1/2^-_1)$ and   $^{13}_\Lambda\textrm{C}(2^+)$, 
%the experimental values of $E_x(1^-)$ and $E_x(3/2^+)$ states are used, respectively.
%For $^{12}_\Lambda\textrm{B,C}(3/2^-_2)$, the experimental values of $E_x(^{12}_\Lambda\textrm{B};1^-)$ and $E_x(^{12}_\Lambda\textrm{C};2^-)$ are averaged
%by assuming that Coulomb shift in $^{12}_\Lambda\textrm{B}(3/2^-_2)$-$^{12}_\Lambda\textrm{C}(3/2^-_2)$ is the same value as that in 
%$^{11}\textrm{B}(3/2^-_2)$-$^{11}\textrm{C}(3/2^-_2)$. 
\label{tab:L-nuclei-ex} }
\begin{center}
\begin{tabular}{ccccccccccccc}
\hline
	&	$r_\Lambda$ &	$\langle k_f \rangle_\Lambda$	&	$R_N$	&$\delta_\Lambda({R_N})$	&  $R_{N}-R_{N,\textrm{gs}}$ 
& $^{A-1}E_x$  &$^{A-1}E_{x,\textrm{exp}}$&$^{A}_\Lambda E_x$  &
$^{A}_\Lambda E_{x,\textrm{exp}}$  & 
$\delta_\Lambda({E_x})$	&	$\delta_\Lambda({E_{x}})_\textrm{DD}$	& 	 $\delta_\Lambda({E_{x}})_\textrm{exp}$	\\
$^7\textrm{Li}(3^+)$	&	2.42 	&	1.02 	&	2.04 	&$	-0.41 	$&$	-0.19 	$&	2.08 	&	2.19 	&	0.89 	&	1.86 	&	-1.19 	&	-0.21 	&	$-0.33$	\\
$^9\textrm{Be}(2^+)$	&	2.41 	&	0.99 	&	2.42 	&$	-3.39 	$&$	-0.02 	$&	$3.11^{\textrm{RGM}}$	&	3.04 	&	2.68 	&	3.04 	&	$-0.43^{\textrm{RGM}}$ 	&	$-0.29^{\textrm{RGM}}$	&	0	\\
$^{11}\textrm{B}(1^+)$	&	2.50 	&	1.03 	&	2.47 	&$	-0.12 	$&$	0.18 	$&	1.21 	&	0.72 	&	2.72 	&	1.67 	&	1.51 	&	0.23 	&	0.95	\\
$^{12}\textrm{B}(1/2^-_1)$	&	2.44 	&	1.09 	&	2.40 	&$	-0.10 	$&$	0.16 	$&	2.79 	&	2.13 	&	4.13 	&	3.00 	&	1.34 	&	0.02 	&	0.87 $(1^-)$	\\
$^{12}\textrm{B}(3/2^-_2)$	&	2.48 	&	1.06 	&	2.46 	&$	-0.12 	$&$	0.22 	$&	5.57 	&	5.02 	&	7.45 	&	6.02 	&	1.88 	&	0.11 	&	1.00*	\\
$^{12}\textrm{B}(5/2^-_1)$	&	2.44 	&	1.09 	&	2.40 	&$	-0.11 	$&$	0.16 	$&	4.66 	&	4.45 	&	6.05 	&		&	1.39 	&	0.03 	&		\\
$^{12}\textrm{C}(1/2^-_1)$	&	2.43 	&	1.09 	&	2.41 	&$	-0.11 	$&$	0.16 	$&	2.62 	&	2.00 	&	4.01 	&	2.73 	&	1.39 	&	0.04 	&	0.73 $(1^-)$	\\
$^{12}\textrm{C}(3/2^-_2)$	&	2.48 	&	1.06 	&	2.48 	&$	-0.12 	$&$	0.23 	$&	5.35 	&	4.80 	&	7.30 	&	5.81 	&	1.96 	&	0.13 	&	1.01*	\\
$^{12}\textrm{C}(5/2^-_1)$	&	2.43 	&	1.09 	&	2.41 	&$	-0.11 	$&$	0.16 	$&	4.50 	&	4.32 	&	5.94 	&		&	1.44 	&	0.05 	&		\\
$^{13}\textrm{C}(2^+)$	&	2.44 	&	1.12 	&	2.39 	&$	-0.10 	$&$	0.13 	$&	4.47 	&	4.44 	&	5.50 	&	4.89 	&	1.03 	&	$-0.04 $	&	0.45 $(3/2^+)$	\\
w/o $p_{3/2}$	&	2.44 	&	1.12 	&	2.39 	&$	-0.10 	$&$	-0.02 	$&	2.36 	&	4.44 	&	2.19 	&	4.89 	&	$-0.17$ 	&	$-0.01$ 	&	0.45 $(3/2^+)$	\\
\hline		
\end{tabular}
\end{center}
\end{table*}

\begin{table}[ht]
\caption{ $B(E2)$  (e$^2$fm$^4$) for $I^\pi_i\to \textrm{g.s.}$ in $^{A-1}Z$ and 
 $^{A}_\Lambda Z$.
For $^{A}_\Lambda Z$, the E2 transition strengths in the core nuclear part $B(E2,\textrm{core})$ are shown. 
The reduction factor $S_{E2}$  is 
also shown. The experimental $B(E2;\textrm{core})$ for $^7_\Lambda \textrm{Li}$ is evaluated 
from the experimental $B(E2;5/2^+\to 1/2^+)$  \cite{Tanida:2000zs} 
by scaling the spin factor $9/7$ as $B(E2;3^+\to 1^+,\textrm{core})=(9/7)B(E2;5/2^+\to 1/2^+)$.
The experimental data for $^{A-1}Z$ nuclei are from Refs.~\cite{AjzenbergSelove:1990zh,Tilley:2002vg,Tilley:2004zz,Kelley:2012qua}.
\label{tab:be2-L} }
\begin{center}
\begin{tabular}{cccccccc}
\hline
 $^{A-1}{Z}(I^\pi_i)$ &\multicolumn{2}{c}{$B(E2)$} &$^{A-1}{Z}(I^\pi_i)$ & \multicolumn{2}{c}{$B(E2,\textrm{core})$} &
\multicolumn{2}{c}{ $S_{E2}$} \\ 
&  cal  & exp  && cal & exp & cal & exp \\
$^6\textrm{Li}(3^+_1)$	&	11.3 	&	10.7(8)	&$^7_\Lambda\textrm{Li}(3^+_1)$	& 3.4 	&4.6(1.3)  &	0.74  & 	0.81(4) \\
$^8\textrm{Be}(2^+_1)$	&		&		&	$^9_\Lambda\textrm{Be}(2^+_1)$&     15.2 	&	&&	\\
$^{10}\textrm{B}(1^+_1)$	&	5.2 	&	4.15(2)	&$^{11}_\Lambda\textrm{B}(1^+_1)$	& 3.1 	&	&0.88& 	\\
$^{11}\textrm{B}(5/2^-_1)$	&	9.5 	&	8.9(3.2)	&$^{12}_\Lambda\textrm{B}(5/2^-_1)$	& 4.5 	&	&0.83& 	\\
$^{12}\textrm{C}(2^+_1)$	&	7.3 	&	7.6(4)	&	$^{13}_\Lambda\textrm{C}(2^+_1)$& 5.0 	&	&0.91& 	\\
w/o $p_{3/2}$ 	&	10.6 	&		&	& 7.8 	& 	&0.93& 	\\
\hline		
\end{tabular}
\end{center}
\end{table}

\subsubsection{Sizes and $E2$ strengths}

In Table \ref{tab:L-nuclei-ex},
the calculated values of the $\Lambda$ distribution size $r_\Lambda$, averaged Fermi momentum $\langle k_f \rangle_\Lambda$, core nuclear size $R_N$, and  nuclear size change $\delta_\Lambda(R_N)$ for excited states obtained with ESC08a(DI)  are shown.
The nuclear size change $\delta_\Lambda(R_N)$ for excited states shows similar trend to that for the ground states. Namely, slight reduction of the nuclear size occurs in $^A_\Lambda Z$ for $A>10$.  
In $^7_\Lambda\textrm{Li}$, 
significant size  reduction occurs also in the excited state, $^7_\Lambda\textrm{Li}(3^+_1)$, because of the spatially developed $\alpha+d$ clustering. 
$^8\textrm{Be}(2^+_1)$  is a broad resonance, but it is bound in  $^9_\Lambda\textrm{Be}(2^+_1)$ because of the $\Lambda$ attraction.

Table \ref{tab:be2-L} shows the $E2$ transition strengths calculated with ESC08a(DI). 
The $B(E2;I^\pm_i\to I^\pm_f,\textrm{core})$ of the 
core nuclear part in $^A_\Lambda Z$ are shown compared with the original $B(E2)$ in $^{A-1}Z$ systems without the $\Lambda$ particle.
$B(E2,\textrm{core})$ in $^A_\Lambda Z$ is generally smaller than the original $B(E2)$ in  $^{A-1}Z$
because of the nuclear size reduction.
We also show the size reduction factor $S_{E2}$ reduced from the ratio of $B(E2;I^\pm_i\to I^\pm_f,\textrm{core})$ in  $^A_\Lambda Z$ 
to the unperturbative value, $B(E2;I^\pm_i\to I^\pm_f)$ in  $^{A-1}Z$, as 
$S_{E2}=[B(E2;I^\pm_i\to I^\pm_f,\textrm{core})/B(E2;I^\pm_i\to I^\pm_f)]^{1/4}$.
In $A>10$ systems, the $B(E2)$ reduction is not as remarkable as that in $A<10$ systems 
because of the small size reduction in the ground and excited states. 
By contrast, $B(E2)$ is remarkably reduced in $^7_\Lambda\textrm{Li}$ as a result of the significant size reduction in the ground and excited states.
This is nothing but the famous phenomenon of the so-called ``glue-like role'' 
of the $\Lambda$ particle \cite{Motoba:1984ri,motoba85}.
% experimentally evidenced for $^7_\Lambda\textrm{Li}$ \cite{Tanida:2000zs}. 
The calculated $B(E2;3^+_1 \to 1^+_1,\textrm{core})$ in $^7_\Lambda\textrm{Li}$ and $B(E2;3^+_1 \to 1^+_1)$ in $^6\textrm{Li}$ 
agree with the experimental data. Detailed discussions are given later. 
%originally predicted in Refs.~\cite{Motoba:1984ri,motoba85}.
%, which has been predicted and discussed by Motoba and his co-workers
%for various cluster systems %(see Ref.~\cite{Hiyama:2010zzc} and references therein)
%\cite{Motoba:1984ri,motoba85,Yamada:1984ii,Sakuda:1987pk,Hiyama:1997ub,Hiyama:1999me, Hiyama:2002yj,Hiyama:2010zzc} 
%and experimentally evidenced for $^7_\Lambda\textrm{Li}$
%by the precise $\gamma$-ray measurement \cite{Tanida:2000zs}. 

\subsubsection{Excitation energy and size difference}

%%%%%%%%%%%%%%%%%%%%%%%%%%%%%%
\begin{figure*}[!h]
\begin{center}
\includegraphics[width=16.0cm]{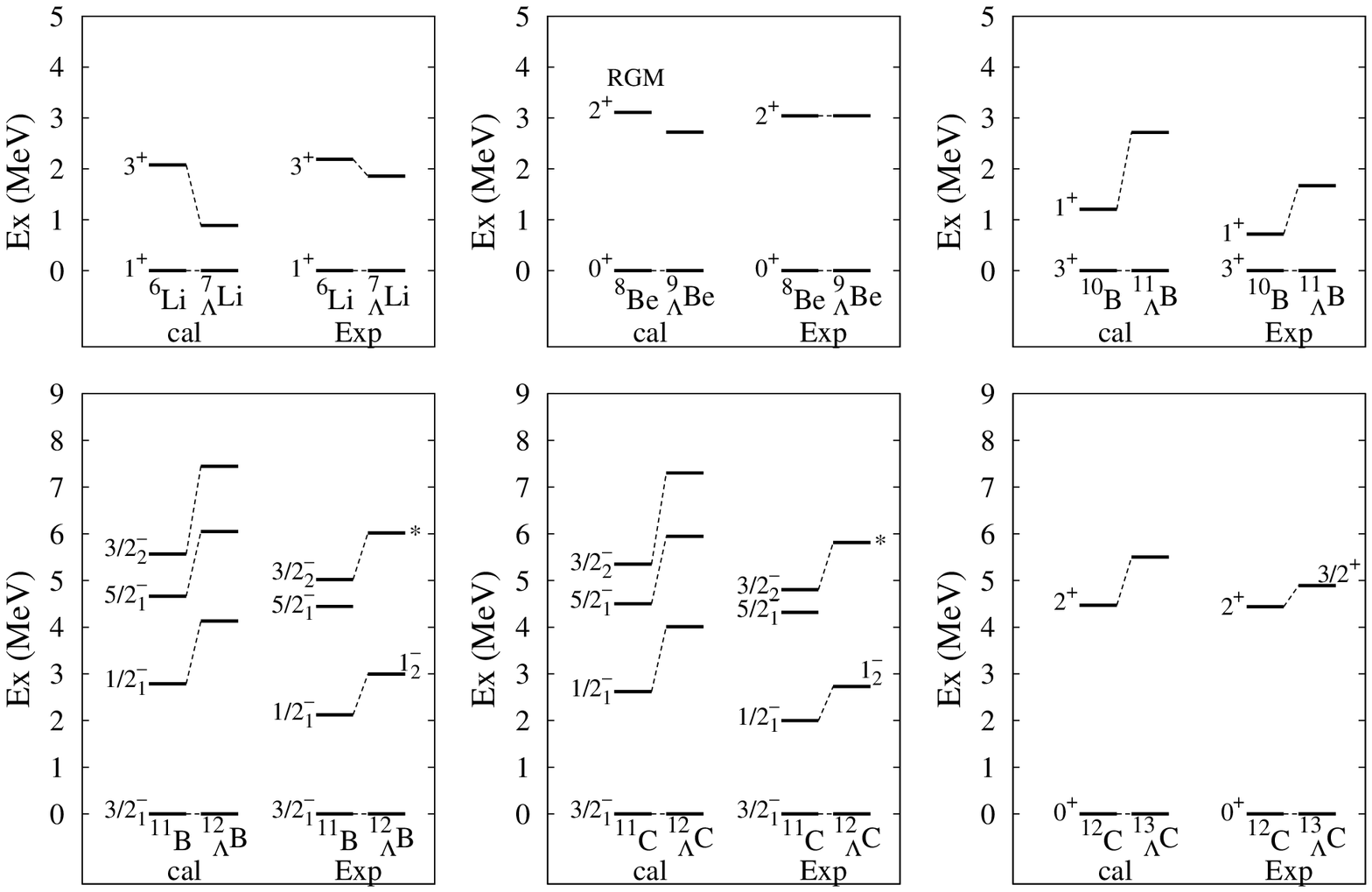} 	
\end{center}
%\vspace{0.5cm}
  \caption{Energy spectra calculated with 
ESC08a(DI) and the experimental spectra.
The experimental data for $^{A-1}Z$ are from Refs.~\cite{Tilley:2002vg,Tilley:2004zz,AjzenbergSelove:1990zh,Kelley:2012qua},
and those for  $^A_\Lambda Z$ are taken 
from Refs.~\cite{Tamura:2010zz,Tanida:2000zs,Ukai:2006zp,Akikawa:2002tm,Miura:2005mh,Ma:2010zzb,Ajimura:2001na,Kohri:2001nc,Tang:2014atx,Hosomi:2015fma}.
The excitation energy of $^8\textrm{Be}(2^+)$ is calculated by the RGM calculation. 
The experimental data for $^A_\Lambda Z$ are the spin-averaged values reduced from  the excitation energies of spin partners, $J^\pi=I^\pi \pm 1/2$,
 except for $^{12}_\Lambda\textrm{B}(1/2^-_1,3/2^-_2)$,
$^{12}_\Lambda\textrm{C}(1/2^-_1,3/2^-_2)$, and   $^{13}_\Lambda\textrm{C}(2^+)$. For $^{12}_\Lambda\textrm{B}(1/2^-_1)$,
$^{12}_\Lambda\textrm{C}(1/2^-_1)$, and   $^{13}_\Lambda\textrm{C}(2^+)$, 
the experimental values of $E_x(1^-)$, $E_x(1^-)$, and $E_x(3/2^+)$  are used, respectively.
For $^{12}_\Lambda\textrm{B,C}(3/2^-_2)$, the experimental values of $E_x(1^-)$ in  $^{12}_\Lambda\textrm{B}$ and $E_x(2^-)$ in  $^{12}_\Lambda\textrm{C}$ are averaged
by assuming that the Coulomb shift in $^{12}_\Lambda\textrm{B}(3/2^-_2)$-$^{12}_\Lambda\textrm{C}(3/2^-_2)$ is the same value as that in 
$^{11}\textrm{B}(3/2^-_2)$-$^{11}\textrm{C}(3/2^-_2)$. 
%Energies are measured from the ground state energy obtained by the full sAMD+GCM.
\label{fig:spe1}}
\end{figure*}
%%%%%%%%%%%%%%%%%%%%%%%%%%%%%

In Table \ref{tab:be2-L}, excitation energies ($E_x$) in  $^{A-1}Z$ and  $^A_\Lambda Z$ are listed. 
The calculated and experimental energy spectra 
are shown in Fig.~\ref{fig:spe1}.
To see the effects of the $\Lambda$ particle on excitation energies, we also show the excitation energy shift 
$\delta_\Lambda(E_x)=E_x(^A_\Lambda Z(I^\pi))-E_x(^{A-1}Z(I^\pi))$ in Table \ref{tab:be2-L}.
%For the experimental data of  $^A_\Lambda Z$, $(2J+1)$-weighted averaged values of spin partners 
%$J=I\pm 1/2$ are shown except for 
%$^{12}_\Lambda\textrm{B}(1/2^-_1,3/2^-_2)$,
%$^{12}_\Lambda\textrm{C}(1/2^-_1,3/2^-_2)$, and   $^{13}_\Lambda\textrm{C}(2^+_1)$. For $^{12}_\Lambda\textrm{B}(1/2^-_1)$ and  
%$^{12}_\Lambda\textrm{C}(1/2^-_1)$, the  experimental values of $E_x(1^-)$ are shown, and 
%for $^{13}_\Lambda\textrm{C}(2^+_1)$,  the data of $E_x(3/2^+)$ is shown.  
%For $^{12}_\Lambda\textrm{B,C}(3/2^-_2)$, the experimental values of $E_x(1^-)$ 
%in $^{12}_\Lambda\textrm{B}$ and $E_x(2^-)$ in $^{12}_\Lambda\textrm{C}$ are averaged
%by assuming the same Coulomb shift in $^{12}_\Lambda\textrm{B}(3/2^-_2)$-$^{12}_\Lambda\textrm{C}(3/2^-_2)$ as that in 
%$^{11}\textrm{B}(3/2^-_2)$-$^{11}\textrm{C}(3/2^-_2)$. 

The ESC08a(DI) result shows the significant energy shift and qualitatively describes the systematic trend of 
the experimental energy shift.
The energy shift comes from the size difference between the ground and excited states because a  $\Lambda$ particle feels a 
deeper potential in a higher nuclear density system through the $\Lambda$-$N$ interactions. 
As shown in Table \ref{tab:L-nuclei-ex}, the excitation energy shift $\delta_\Lambda(E_x)$ clearly correlates with 
the size difference $R_N-R_{N,\textrm{gs}}$, where $R_{N,\textrm{gs}}$ is the size of the ground state.
Namely, the excitation energies shift upward reflecting the larger sizes of excited states in $A>10$ systems. 
Note that, in $A>10$ systems, the size difference in $^{A}_\Lambda Z$ is  consistent with that in $^{A-1}Z$
meaning that the origin of the size difference, i.e.,  the excitation energy shift, is the structure difference between the ground 
and excited states in original core nuclei $^{A-1}Z$.
The excited state $^{10}\textrm{B}(1^+_1)$ has a developed $2\alpha+d$ cluster, and has 
a larger size than that of the ground state $^{10}\textrm{B}(3^+_1)$ with a
weaker clustering because of the stronger spin-orbit attraction 
of the $d$ cluster as discussed in Refs.~\cite{Morita:2016yql,Morita:2017gmy}.
The excited states of $^{11}\textrm{B}$ and $^{11}\textrm{C}$ have the $2\alpha+t$ and $2\alpha+^3\textrm{He}$ 
cluster structures, and  have larger sizes than those of the ground states $^{11}\textrm{B}(3/2^+_1)$ and  $^{11}\textrm{C}(3/2^+_1)$, respectively, 
which are reduced by the cluster breaking ($p_{3/2}$) component.
Also in $^{12}\textrm{C}$, the excited state 
$^{12}\textrm{C}(2^+_1)$ has the $3\alpha$ cluster structure and the larger size than that of the ground state, in which  significant 
mixing of the cluster breaking component reduces the size of the ground state.
In $^7_\Lambda\textrm{Li}$, the situation is opposite. The excited states, $^6_\Lambda\textrm{Li}(3^+_1)$ and 
$^7_\Lambda\textrm{Li}(3^+_1)$ have smaller sizes than those of the ground states, $^6_\Lambda\textrm{Li}(1^+_1)$ and 
$^7_\Lambda\textrm{Li}(1^+_1)$, 
because of the higher centrifugal barrier in addition to the stronger spin-orbit attraction between 
$\alpha$ and $d$ clusters  in the $D$-wave $\alpha+d$ state than in the $S$-wave state.
Reflecting the smaller size than the ground state, the excitation energy of $^7_\Lambda\textrm{Li}(3^+_1)$ shifts downward.
For $^9_\Lambda\textrm{Be}$,  it is difficult to give a quantitative discussion of the energy shift because $^8\textrm{Be}(2^+)$ is the broad resonance.

For $^{13}_\Lambda\textrm{C}$, the traditional $3\alpha$ calculation without the cluster breaking  gives a result different 
from the present result obtained by the $3\alpha+p_{3/2}$ calculation. 
In the traditional $3\alpha$ calculation, the ground state has the $3\alpha$ cluster structure with no cluster breaking 
and almost the same or even slightly larger size than the excited state.  The comparable sizes between the ground and excited states are reflected in 
the small excitation energy shift in $^{13}_\Lambda\textrm{C}(2^+_1)$ in the traditional $3\alpha$ calculation.
This contradicts to the present result and is inconsistent with the
experimental data.  It should be commented that 
traditional 3$\alpha$-cluster models generally obtain a slightly smaller size of $^{12}\textrm{C}(2^+_1)$ than the ground state.
For example, the size of 
$^{12}\textrm{C}(2^+_1)$ and that of $^{12}\textrm{C}(0^+_1)$ are 2.38 fm and 2.40 fm  in the $3\alpha$ RGM calculation\cite{kamimura-RGM2} 
(2.50 fm and 2.53 fm in the 3$\alpha$ GCM calculation \cite{uegaki1,uegaki3}). It means that cluster breaking components in core nuclei can affect 
the excitation energy shift in $^A_\Lambda Z$ systems.

%%%%%%%%%%%%%%%%%%%%%%%%%%%%%%
\begin{figure}[!h]
\begin{center}
\includegraphics[width=8.0cm]{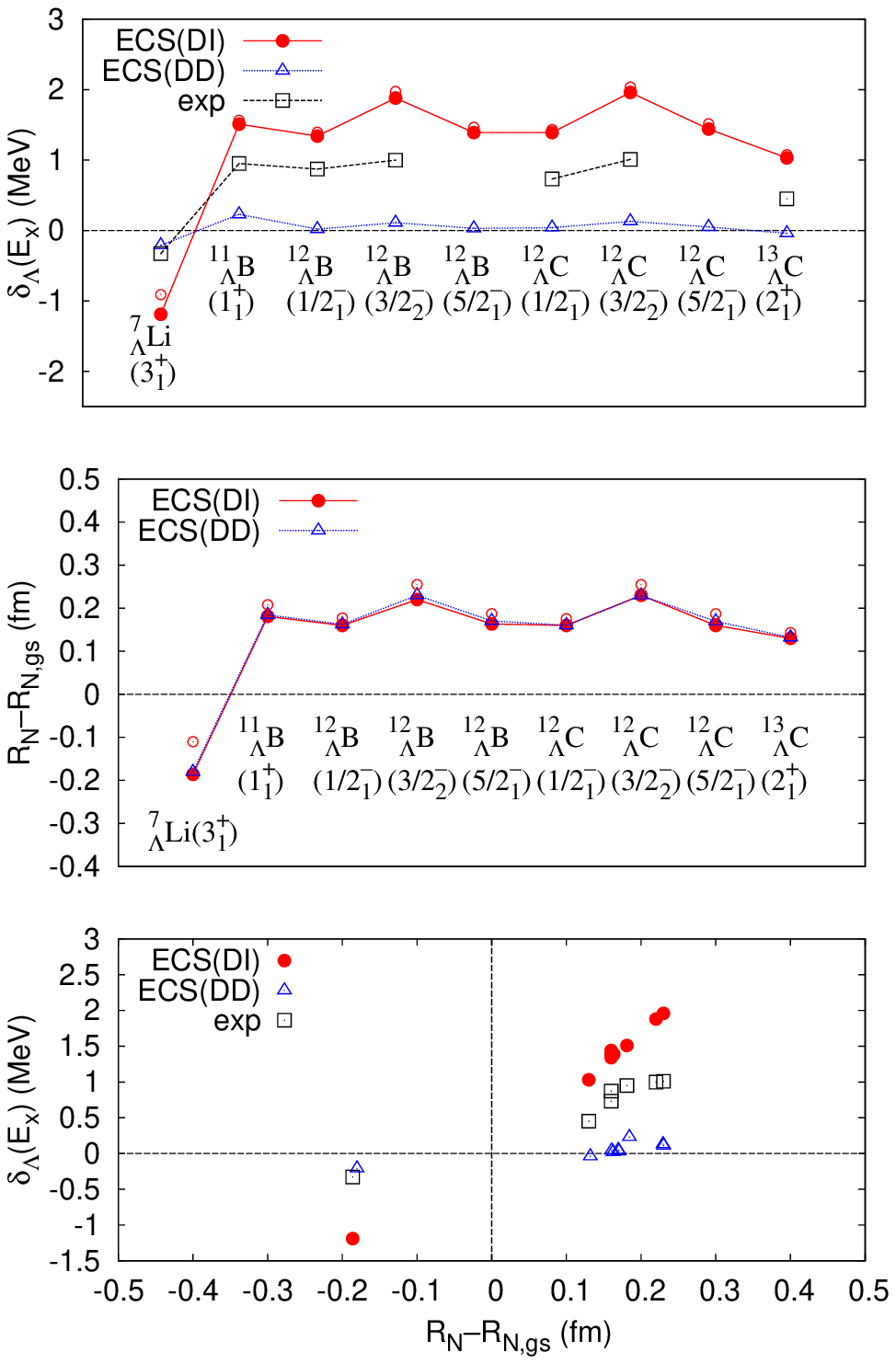} 	
\end{center}
%\vspace{0.5cm}
  \caption{(color online) 
(top) Excitation energy shift $\delta_\Lambda(E_x)$
calculated with ESC08a(DI) and  ESC08a(DD), and experimental values. 
The excitation energy shift without the core polarization calculated with ESC08a(DI) is also plotted by
open circles. (middle) Nuclear size difference $(R_N-R_{N,\textrm{gs}})$ 
of excited states from that of the ground states obtained with 
 ESC08a(DI) and  ESC08a(DD).  The result without the core polarization calculated with ESC08a(DI) is also shown  by
open circles. (bottom)  The excitation energy shift plotted against the nuclear size difference obtained with 
 ESC08a(DI) and  ESC08a(DD). Experimental energy shift is 
plotted against the theoretical size difference
calculated with ESC08a(DI).
%Energies are measured from the ground state energy obtained by the full sAMD+GCM.
\label{fig:r-e-corr}}
\end{figure}
%%%%%%%%%%%%%%%%%%%%%%%%%%%%%

In order to look into the dependence of the excitation energy shift on the size difference in more detail, 
we plot the energy shift and the size difference 
in Fig. \ref{fig:r-e-corr}. The ESC08a(DI)  results show a  clear correlation between the energy shift and the size difference.
Namely, the larger size, the larger energy shift. 
%The calculated and experimental energy shifts are plotted against the calculated size difference. 
The calculation qualitatively describes the systematic trend of the experimental data. However, quantitatively, it 
overestimates the experimental energy shift by a factor of $1.5-2$. 

Although it is generally difficult to experimentally measure 
sizes of excited states in $^{A-1}Z$ systems,  
we can obtain information from the Coulomb shift in mirror nuclei. 
As described previously,
the present calculation reasonably reproduces 
the experimental Coulomb shift in $^{11}\textrm{B}$-$^{11}\textrm{C}$.
 Roughly speaking, about 0.2 fm size difference describes $\sim 0.2$ MeV Coulomb shift
in $^{11}\textrm{B}$-$^{11}\textrm{C}$, whereas it gives $\sim$1 MeV energy shift 
in $^{12}_\Lambda\textrm{B}$ ($^{12}_\Lambda\textrm{C}$). Namely,   the size difference causes 
about $5$ times larger energy difference in the $\Lambda$ energy than that in the Coulomb energy.

The energy shift and size difference calculated without the core polarization are also shown in Fig.~\ref{fig:r-e-corr} by open circles.
They are almost consistent with the results with the core polarization because the core polarization (size reduction) effect on energy is higher-order perturbation.
In other words, the origin of the excitation energy shift in $^A_\Lambda Z$ is, in the leading order, 
the nuclear size difference between the ground and excited states in the original (unperturbative) core nuclei $^{A-1}Z$. 
It turns out that the excitation energy shift in $\Lambda$ hypernuclei can  probe 
the size difference between the ground and excited states in original $^{A-1}Z$ nuclei in this mass-number region.

In contrast to the significant energy shift in the ESC08a(DI) results, the ESC08a(DD) results show almost no energy shift
and fails to describe the systematic trend of the experimental energy shift as shown in Table \ref{tab:L-nuclei-ex} and Fig.~\ref{fig:r-e-corr}. 
In the case of ESC08a(DD),
the $\Lambda$ energy in $^A_\Lambda Z$  has no (or only weak) dependence on the nuclear size 
because of the density ($k_f$) dependence of the $\Lambda NG$ interactions as discussed previously, and therefore, the $\Lambda$ particle can not probe the
nuclear size difference between the ground and excited states. 
%In recent study of hypernuclei with the AMD in Refs.~\cite{}, 
%the averaged nuclear density $\bar\rho_N$ is used 
%for $k_f=(3\pi^2\bar\rho_N/2)^{1/3}$  in the density-($k_f$-)dependence of the $\Lambda NG$ interactions  instead of the LDA. 
%The calculation using
%the self-consistently determined $\bar\rho_N/2$ gives similar result to  the ESC08a(DD) because 
%the density dependence of the $\Lambda-N$ interactions generally make the sensitivity 
%of $E_\Lambda$ to the nuclear size as weak as ESC08a(DD).

As seen in the bottom panel of Fig.~\ref{fig:r-e-corr}, 
 the systematic trend of the experimental energy shift can be described  by the ESC08a(DI) calculation
but not by the ESC08a(DD) calculation, meaning that the density-independent $\Lambda NG$ interactions 
are rather favored than the density-dependent ones. However, as for the quantitative reproduction, 
the ESC08a(DI) calculation generally overestimates the experimental energy shift by a factor of $1.5-2$. 
It is likely that weak density dependence of the $\Lambda NG$ interactions 
may be suitable 
for detailed description of the excitation energy shift  in $^A_\Lambda Z$.

\section{Discussions}\label{sec:discussions}

\subsection{Density distributions}

%%%%%%%%%%%%%%%%%%%%%%%%%%%%%%
\begin{figure*}[!h]
\begin{center}
\includegraphics[width=15.0cm]{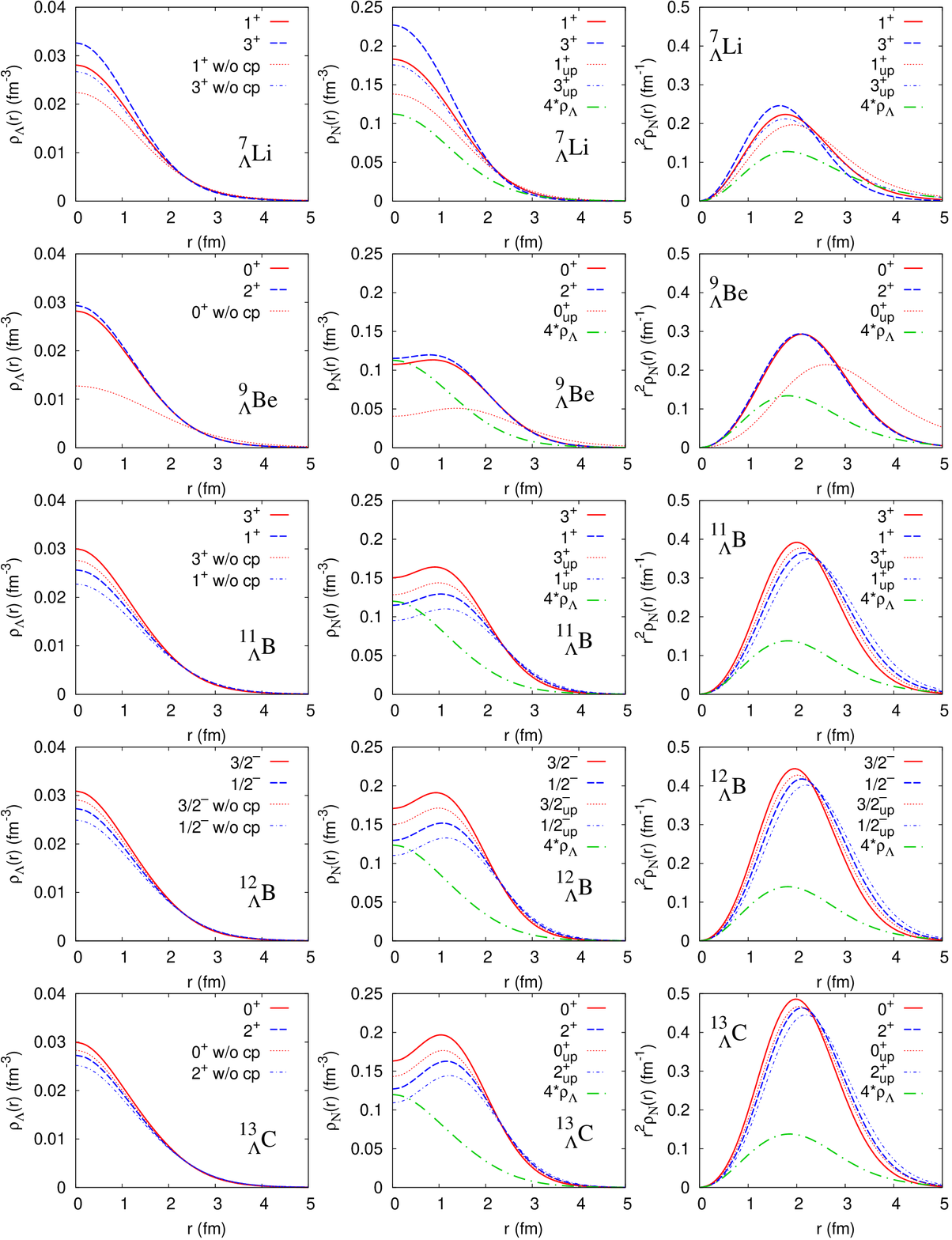} 	
\end{center}
%\vspace{0.5cm}
  \caption{(color online) $\Lambda$ density and nuclear density distributions
 in 
$^7_\Lambda\textrm{Li}$, $^{9}_\Lambda\textrm{Be}$, $^{11}_\Lambda\textrm{B}$, 
$^{12}_\Lambda\textrm{B}$, and 
$^{13}_\Lambda\textrm{C}$ calculated with ESC08a(DI). 
In the left panels, 
$\Lambda$ density 
obtained in the calculation with and without the core polarization (w/o cp) are shown. 
In the middle and right panels, nuclear densities and 
$r^2$-weighted densities are shown, respectively.  
In the middle panels, 
$\Lambda$ density ($\rho_\Lambda(r))$ multiplied by 4 in the ground state of $A_\Lambda Z$ is also shown
for comparison. 
\label{fig:rho-N}}
\end{figure*}
%%%%%%%%%%%%%%%%%%%%%%%%%%%%%
Figure \ref{fig:rho-N} shows the distribution functions of 
the $\Lambda$ density $\rho_\Lambda(r)$  and nuclear density $\rho_N(r)$
in the ground and excited states of $^A_\Lambda Z$ 
as functions of $r$. Note  that $r$ is  
the distance from the cm of  core nuclei. 
The figures also show the $\Lambda$ density $\rho^\textrm{up}_\Lambda(r)$ and nuclear density  $\rho^\textrm{up}_N(r)$
in $\Lambda\textrm{-}(^{A-1}Z)$ systems 
with unperturbative core nuclei (without core polarization). 
$\rho^\textrm{up}_N(r)$ is the original nuclear density in isolate $^{A-1} Z$ systems
without the $\Lambda$ particle. 

Let us discuss the $\Lambda$ density shown in the left panels of Fig.~\ref{fig:rho-N}.
As seen in the small difference between $\rho_\Lambda(r)$ and $\rho^\textrm{up}_\Lambda(r)$, 
the core polarization effect on the $\Lambda$ distribution 
is rather minor except for $^9_\Lambda \textrm{Be}$. 
Moreover, the difference between the ground and excited states
in each system is also small. The center $\Lambda$ density increases 
as the mass number $A$ increases reflecting the deeper $\Lambda$ binding in heavier systems.

Let us look at the nuclear density shown in the middle and right panels of  Fig.~\ref{fig:rho-N}.  
%The core polarization effect can be seen in the nuclear density.  
Compared with the original density  $\rho^\textrm{up}_N(r)$ in  $^{A-1} Z$ systems, 
the nuclear density $\rho_N(r)$ in  $^{A}_\Lambda Z$
is slightly  increased
in $A>10$ systems, and rather significantly enhanced in $A<10$ systems,
as the result of the size reduction discussed previously.

In comparison of the nuclear density between the ground and excited states in each system, 
we find significant difference between them except for $^9_\Lambda\textrm{Be}$.
 In $A>10$ systems, the inner density (typically in the $r \lesssim 2$ region) 
is lower in excited states
than in the ground states. 
The situation is opposite in $^7_\Lambda\textrm{Li}$. The $^7_\Lambda\textrm{Li}(3^+_1)$ 
has the higher inner density than that of $^7_\Lambda\textrm{Li}(1^+_1)$.
As shown by green lines in right panels of Fig.~\ref{fig:rho-N}, the $r^2$-weighted $\rho_\Lambda(r)$ has a maximum peak 
at $r=1.5\sim 2.0$ fm, and therefore, the $\Lambda$ particle mainly probes the density difference between the ground 
and excited states in this region. Reflecting the nuclear density (size) difference between the ground and excited states,
the excitation energy shift occurs. The trend is similar in the results with and without the core polarization
because the core polarization slightly raises the inner nuclear density with almost the same amount 
in both ground and excited states.
%Since the peak position of the $r^2$-weighted $\rho_\Lambda(r)$ shifts inward  as $A$ increases, 
%because the $\Lambda$ particle is bound more deeply inside the nuclei.  Therefore, it is likely that 
%the $\Lambda$ particle in heavier systems can probe the density difference in smaller $r$ regions. 
%It should be commented that $^9_\Lambda\textrm{Be}$ has the inner nuclear density lower 
%than the normal density because it still has a loosely bound $2\alpha$-cluster nature even though the 
%inner nuclear density is increased by the $\Lambda$ attraction. 
%It should be also mentioned that $^7_\Lambda\textrm{Li}$ has the rather high nuclear density at the center 
%because of much weaker Pauli blocking between $\alpha$ and $d$ clusters than that between 2 $\alpha$ clusters. 

\subsection{Size shrinkage in $^7_\Lambda\textrm{Li}$ and $^9_\Lambda\textrm{Be}$: comparison with other 
cluster model calculations}

%The size shrinkage of cluster structure is one of the typical phenomena arising from the glue-like role of 
%$\Lambda$ particle originally predicted by Motoba {\it et. al.} \cite{Motoba:1984ri, motoba85}.
The shrinkage of cluster structures in  $^7_\Lambda\textrm{Li}$ and $^9_\Lambda\textrm{Be}$ has been 
theoretically investigated in connection with $\gamma$ transitions in details 
with the semi-microscopic $\alpha+d+\Lambda$ and $2\alpha+\Lambda$ cluster models, respectively,  
using the OCM \cite{Motoba:1984ri,motoba85}. 
The shrinkage and $\gamma$ transitions  in $^7_\Lambda\textrm{Li}$ have  been also investigated 
with the semi-microscopic $^5_\Lambda\textrm{He}+p+n$ cluster OCM calculation \cite{Hiyama:1999me}.
The predicted size shrinkage in $^7_\Lambda\textrm{Li}$ has been evidenced by the experimental
measurement of the $E2$ transition strengths for the $5/2^+\to 1/2^+$ transition \cite{Tanida:2000zs}. 

We here describe the results for the shrinkage properties 
in $^7_\Lambda\textrm{Li}$ and $^9_\Lambda\textrm{Be}$, 
and compare them with results of Refs.~\cite{Motoba:1984ri,Hiyama:1999me}.
We also show comparison with the recent calculation of  $^9_\Lambda\textrm{Be}$ with a microscopic $2\alpha+\Lambda$ cluster 
model with the $S$-wave $\Lambda$ assumption in Ref.~\cite{Funaki:2014fba}.
The interactions for the $\Lambda$ particle in Ref.~\cite{Motoba:1984ri} are phenomenological $\Lambda$-cluster potentials, and 
those in Ref.~\cite{Hiyama:1999me} are the $\Lambda$-cluster  potentials derived from the $\Lambda NG$ interactions with 
phenomenologically adjusted $k_f$ parameters.
In Ref.~\cite{Funaki:2014fba},  the density-independent $\Lambda NG$ interactions with  a fixed $k_f$ parameter 
are used as the effective $\Lambda$-$N$  interactions.
These effective interactions are state-independent (structure-independent) and, in that sense, they correspond to the 
density-independent treatment of $k_f$ in ECS08a(DI) in the present calculation. 
For the results of Refs.~\cite{Hiyama:1999me,Funaki:2014fba}, 
we show the values of the Nijmegen type-D (ND) case 
of the $\Lambda NG$  parrametrization. 

In the works with semi-microscopic cluster models, the size shrinkage, i.e., the contraction of the $\alpha+t$ and $\alpha+\alpha$ cluster structures 
in $^7_\Lambda\textrm{Li}$  and $^9_\Lambda \textrm{Be}$ is usually discussed for the reduction of the rms distance 
between clusters because they are directly related to electric transition strengths in two-body cluster states. 
In particular, the $E2$ transition strength from the  $D$-wave excited state to the $S$-wave ground state is 
sensitive to the shrinkage because it is approximately proportional to the fourth power of the inter-cluster distance.
To discuss the shrinkage of the cluster structures and its relation to the $E2$ transitions, 
we approximately estimate the rms inter-cluster distances $\overline{r}_{\alpha\textrm{-}x}\equiv \langle r^2_{\alpha\textrm{-}x} \rangle^{1/2}$  
between $\alpha$ and $x$ clusters from the
calculated nuclear matter radius $R_N$ using the following simple relation for non-microscopic two clusters, 
\begin{equation}\label{eq:r-app}
(4+A_x) R^2_N=\frac{4 A_x}{4+A_x}\langle r^2_{\alpha\textrm{-}x}\rangle +4 R^2_\alpha + A_x R^2_x, 
\end{equation}
where $x$ is $d$($\alpha$) for $^7_\Lambda\textrm{Li}$($^9_\Lambda \textrm{Be}$), and 
$A_x$ and $R_x$ are the mass number and rms matter radius of the $x$ cluster, respectively.  
We use the theoretical values $R_\alpha=1.55$ fm and  $R_d=1.26$ fm for the $(0s)^4$ and $(0s)^2$ states with  
the present parametrization $\nu=0.235$ fm$^{-2}$. 
We also approximate the $\alpha$-$\alpha$ distance $\overline{r}_{\alpha\textrm{-}\alpha}$ in 
$^9_\Lambda\textrm{Be}$ from $R_N$ in Ref.~\cite{Funaki:2014fba} 
with \eqref{eq:r-app}
using their parameter $\nu=1/(2\cdot 1.36^2)$  fm$^{-2}$.

The size shrinkage in  $^7_\Lambda\textrm{Li}$ is characterized by the reduction of the distance 
$\overline{r}_{\alpha\textrm{-}d}$ from 
$^6\textrm{Li}$ to $^7_\Lambda\textrm{Li}$, and  is discussed with the size reduction factor
\begin{equation}
S= \frac{\overline{r}_{\alpha\textrm{-}d}(^7_\Lambda\textrm{Li})}{\overline{r}_{\alpha\textrm{-}d}(^6\textrm{Li})}.
\end{equation}
The reduction factor can be also reduced from the $E2$ transition strengths for 
$^6\textrm{Li}(3^+_1)\to^6\textrm{Li}(1^+_1)$ and $^7_\Lambda\textrm{Li}(5/2^+)\to ^7_\Lambda\textrm{Li}(1/2^+)$, 
as 
\begin{equation}
S_{E2}= \left[
 \frac{B(E2; ^6\textrm{Li}(3^+_1)\to^6\textrm{Li}(1^+_1))}{(9/7) B(E2;^7_\Lambda\textrm{Li}(5/2^+)\to ^7_\Lambda\textrm{Li}(1/2^+))}
\right ]^{1/4}.
\end{equation}
Here the denominator corresponds to the $E2$ transition strength, $B(E2;I^\pi_i \to I^\pi_f,\textrm{core})$, 
for the $3^+\to 1^+$ transition of the core nuclear part in 
$^7_\Lambda\textrm{Li}$. The factor 9/7 is derived in the weak coupling limit of  the core spin $I$ and the $\Lambda$ intrinsic spin 
\cite{Motoba:1984ri}.

\begin{table}[ht]
\caption{$\alpha$-$d$ distance $\overline{r}_{\alpha\textrm{-} d}$
 in  $^6\textrm{Li}$ and $^7_\Lambda\textrm{Li}$, and 
$\Lambda$ binding energy ($B_\Lambda$) and  $\Lambda$ distribution size ($r_\Lambda$) in $^7_\Lambda\textrm{Li}$ calculated with 
ESC08a(DI) and ESC08a(DD).  The calculated $B(E2;3^+_1\to 1^+_1)$ in $^6\textrm{Li}$, 
$B(E2;3^+_1\to 1^+_1,\textrm{core})$ in $^7_\Lambda\textrm{Li}$, the reduction factors $S$ for the 
$1^+_1$ and $3^+_1$ states, and $S_{E2}$ 
are also listed.
Theoretical values  of other calculations from Refs.~\cite{Motoba:1984ri,Hiyama:1999me}, 
and experimental values from Refs.~\cite{Tilley:2002vg,Davis:2005mb,Tanida:2000zs} 
are also listed.  
%The values from Ref.~\cite{Hiyama:1999me} are those with the ND parametrization of the $\Lambda NG$ interactions.
%$B(E2:3^+\to 1^+,\textrm{core})$ values for Refs.~\cite{Motoba:1984ri,Hiyama:1999me,Tanida:2000zs}
%shown here are the scaled values of (9/7)$B(E2)$ for the $5/2^+\to 1/2^+$ transition  in $^7_\Lambda\textrm{Li}$. 
The $\alpha$-$d$ distance of Ref.~\cite{Hiyama:1999me} is 
the rms $\alpha\textrm{-}(pn)$ distance.
%The experimental $B_\Lambda$ value is  taken from Ref.~\cite{Davis:2005mb}.
\label{tab:shrink-li6} }
\begin{center}
\begin{tabular}{cccccc}
\hline
	&	\cite{Motoba:1984ri}&	\cite{Hiyama:1999me}	&	\multicolumn{2}{c}{present} &	exp	\\
& &  &	DI  & DD & \\
$^6\textrm{Li}$\qquad  & & & & & \\
 $\overline{r}_{\alpha\textrm{-}d}(1^+_1) $(fm)	&	3.8	&	3.85	&	4.45	&	4.45	&		\\
$\overline{r}_{\alpha\textrm{-}d}(3^+_1)$	(fm) &	3.66	&		&	4.16	&	4.16	&		\\
$B(E2)$ ($e^2$fm$^4$)	&	6.6	&	9.62	&	11.3 	&	11.3 	&	10.7(8)	\\
& & & & & \\
$^7_\Lambda\textrm{Li}$\qquad  & & & & & \\
$B_\Lambda$ (MeV)	&	5.59	&	5.58	&	5.44 	&	5.43 	&	5.58(3)	\\
$\overline{r}_{\alpha\textrm{-}d}(1^+_1)$ (fm) &	3.13	&	2.94	&	3.56 	&	4.05 	&		\\
$\overline{r}_{\alpha\textrm{-}d}(3^+_1)$ (fm) &	2.91	&		&	3.02 	&	3.56 	&		\\
$r_\Lambda (1^+_1)$ (fm)	&	2.4	&		&	2.57 	&	2.66	&		\\
$r_\Lambda (3^+_1)$ (fm)	&	2.33	&		&	2.42 	&	2.61	&		\\
$B(E2,\textrm{core})$  ($e^2$fm$^4$) 	&	3.2	&	3.1	&	3.4 	&	6.2 	&	4.6(1.3)	\\
& & & & & \\
$S(1^+_1)$	&	0.82 	&	0.76	&	0.80 	&	0.91 	&	\\
$S(3^+_1)$	&	0.80 	&		&	0.72 	&	0.86 	&	\\
$S_{E2}$	&	0.83 	&	0.75 	&	0.74 	&	0.86 	&	0.81(4)	\\
\hline		
\end{tabular}
\end{center}
\end{table}

\begin{table}[ht]
\caption{ 
$\alpha$-$\alpha$ distance $\overline{r}_{\alpha\textrm{-}\alpha}$
 in  $^8\textrm{Be}$ and $^9_\Lambda\textrm{Be}$, and 
$\Lambda$ binding energy ($B_\Lambda$),  $\Lambda$ distribution size ($r_\Lambda$), and 
$B(E2;2^+\to 0^+,\textrm{core})$ in  $^9_\Lambda\textrm{Be}$ calculated with 
ESC08a(DI) and ESC08a(DD). 
Theoretical values  of other calculations from Refs.~\cite{Motoba:1984ri,Funaki:2014fba}
are also listed.  
%The values from Ref.~\cite{Funaki:2014fba} are those with the ND parametrization of the $\Lambda NG$ interactions.
The experimental $B_\Lambda$ value is from Ref.~\cite{Davis:2005mb}.
\label{tab:shrink-be8} }
\begin{center}
\begin{tabular}{cccccc}
\hline
	&	\cite{Motoba:1984ri}&	\cite{Funaki:2014fba}	&	\multicolumn{2}{c}{present} &	exp	\\
& &  &	DI  & DD & \\
$^8\textrm{Be}\qquad $  & && & & \\
$\overline{r}_{\alpha\textrm{-}\alpha}(0^+_1)$ (fm)	&	4.09	&		4.96	&	5.99 	&	5.99 	&		\\
%Be8(2+)	&	4.17	&		&		&		&		&		\\
& & & & & \\
$^9_\Lambda\textrm{Be}\qquad $ && & & & \\
$B_\Lambda$ (MeV)	&	7.49	&		7.33	&	7.04	&	6.43 	&	6.71(4)	\\
$\overline{r}_{\alpha\textrm{-}\alpha}(0^+_1)$ (fm)	&	3.46	&		3.61	&	3.76 	&	4.41 	&		\\
$\overline{r}_{\alpha\textrm{-}\alpha}(2^+_1)$ (fm) &	3.44	&			3.56	&	3.71 	&	4.65 	&		\\
$r_\Lambda (0^+_1)$ (fm)	&	2.39	&			2.57	&	2.44 	&	2.67	&		\\
$r_\Lambda (2^+_1)$ (fm)	&	2.39	&		2.55	&	2.41 	&	2.67	&		\\
$B(E2,\textrm{core})$ (e$^2$fm$^4$)	&	11.3	&		13.1	&	15.2 	&	31.6 	&		\\
\hline		
\end{tabular}
\end{center}
\end{table}

In Table  \ref{tab:shrink-li6}, we show the calculated results of the distance
$\overline{r}_{\alpha\textrm{-}d}$ in the ground and excited states of 
$^6\textrm{Li}$ and $^7_\Lambda\textrm{Li}$, $B(E2)$ for 
$3^+\to 1^+$, and the reduction factors compared with the theoretical values of 
Refs.~\cite{Motoba:1984ri,Hiyama:1999me}.
For the $B(E2;\textrm{core})$ values, 
the theoretical $B(E2;3^+\to 1^+,\textrm{core})=(9/7)B(E2;5/2^+ \to 1/2^+) $  from Refs.~\cite{Motoba:1984ri,Hiyama:1999me}
and the experimental value are shown. 
In the present calculation with ESC08a(DI), we obtain almost consistent results with those of other calculations. 
The distance $\overline{r}_{\alpha\textrm{-}d}$ is significantly reduced in $^7_\Lambda\textrm{Li}$
from $^6\textrm{Li}$. 
The reduction factor $S_{E2}$ obtained with ESC08a(DI)
agrees with the theoretical values of other calculations, and is in reasonable agreement with 
the experimental value within the error. 
In the ESC08a(DD) result, 
the size shrinkage is relatively small. 

In Table \ref{tab:shrink-be8}, we list the results for $^9_\Lambda\textrm{Be}$ with 
those of other calculations in  Refs.~\cite{Motoba:1984ri,Funaki:2014fba}. 
Also for $^9_\Lambda\textrm{Be}$, 
the present calculation with ESC08a(DI)
gives almost consistent results with those of  Refs.~\cite{Motoba:1984ri,Funaki:2014fba}. 
The significant shrinkage occurs in $^9_\Lambda\textrm{Be}$ as seen in the smaller $\overline{r}_{\alpha\textrm{-}\alpha}$
value than that in $^8\textrm{Be}$. 

\subsection{Interpretation of enhancement factor}

In order to take into account the core polarization in $^A_\Lambda Z$,  
%in the present calculation,
we added the artificial interactions $\Delta H(\epsilon)$  to the Hamiltonian by slightly 
enhancing the central nuclear interactions. In the present cluster models, 
the perturbative interactions, $\Delta H(\epsilon)=\epsilon V^{(c)}_N$, act as slight enhancement of the inter-cluster potentials
between inert clusters. It is consistent with the expectation from the glue-like role of a $\Lambda$ particle.
In a mean-field picture, this treatment corresponds to slight enhancement of the nuclear 
mean potentials $U^{(NN)}_N(r)\to U^{(NN)}_N(r)+\epsilon U^{(NN)}_N(r)$ originating in the $NN$ interactions.
In a self-consistent mean-field approach, nucleons in $^A_\Lambda Z$ feel 
the mean potentials $U^{(NN)}_N(r)+ U^{(\Lambda N)}_N(r)$, where $U^{(\Lambda N)}_N(r)$ is the 
$\Lambda$-$N$-interaction-origin mean potentials for nucleons. 
In the case of $\rho_\Lambda(r)\sim \rho_N(r)/(A-1)$
that the $\Lambda$ distribution function is similar to the nuclear density distribution one, 
$U^{(\Lambda N)}_N(r)$ may be approximated to be 
\begin{equation}
\epsilon U^{(NN)}_N(r) \sim U^{(\Lambda N)}_N(r), 
\end{equation}
which corresponds to the present treatment of the core polarization.
In the present results in $^A_\Lambda Z$ in the $6< A <14$ region, this condition is roughly satisfied 
as seen in the calculated $\Lambda$ and nuclear densities  as well as sizes
$r_\Lambda \sim R_N$.
Considering that $U^{(\Lambda N)}_N(r)\sim U_\Lambda(r)/(A-1)$ in this condition,   
it leads to the relation, 
\begin{eqnarray}
\epsilon U^{(NN)}_N(r) &\sim& U^{(\Lambda N)}_N(r)\sim \frac{1}{A-1} U_\Lambda(r), \\
\epsilon &\sim & \frac {1}{A-1} \frac{U_\Lambda(r)}{U^{(NN)}_N(r)}.
\end{eqnarray}
It means that the enhancement factor $\epsilon$ can be proportional to $1/(A-1)$. 
Figure \ref{fig:r-e-corr} shows the $(A-1)$-dependence of the optimized $\epsilon$ values, which are determined 
for each $^A_\Lambda Z$ system to minimize the total energy.
The $\epsilon$ values are approximately on the $1.2/(A-1)$ line 
except for $^9_\Lambda\textrm{Be}$, for which the mean-field picture may not work well because it is a
dilute $2\alpha$-cluster system.
The additional factor, 1.2,  may come from various origins such as  
possible deviation from the relation $\rho_\Lambda(r)\sim\rho_N(r)/(A-1)$, 
the Pauli blocking between nucleons (no blocking between $\Lambda$ and a
nucleon), 
the weaker $\Lambda$-$N$ interactions than the $N$-$N$ interactions, 
and so on. 
%Note that the dynamical effect of increasing  nuclear interactions on structure change 
%is equivalent to virtually increasing the kinetic mass (except for slight modification of Coulomb interaction term).
%The kinetic mass increase in nuclear systems may cause slight reduction of nuclear size and/or deformation 
% consistently as expected from the additional attraction from the $\Lambda$ particle bound 
%deeply inside nuclei. 

The picture discussed here may support the present treatment of the core polarization at least in the light
mass-number region. However, it may not obvious whether it is useful for heavier systems, in which the 
$\Lambda$ particle is localized deeply inside the core nuclei and
the condition $\rho_\Lambda(r)\sim \rho_N(r)/(A-1)$ is no longer satisfied.

%%%%%%%%%%%%%%%%%%%%%%%%%%%%%%
\begin{figure}[!h]
\begin{center}
\includegraphics[width=6.0cm]{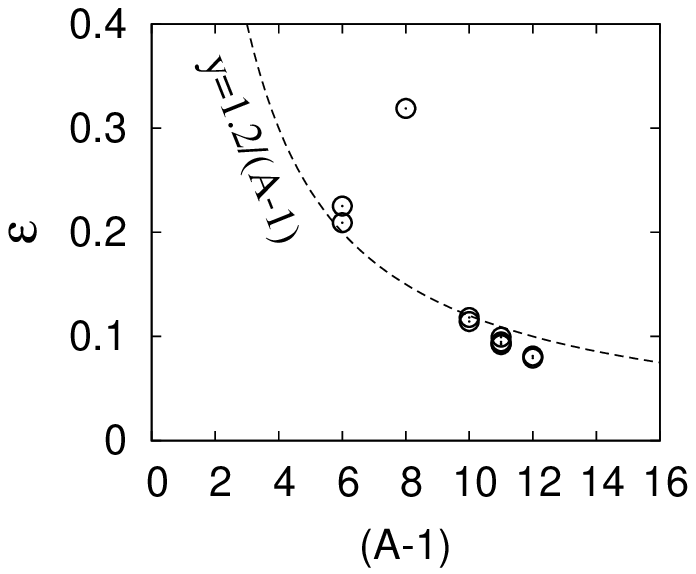} 	
\end{center}
%\vspace{0.5cm}
  \caption{Enhancement factor $\epsilon$ in the ESC08a(DI) calculation. The factor is 
plotted as a function of $A-1$. 
%Energies are measured from the ground state energy obtained by the full sAMD+GCM.
\label{fig:core-pol-fig}}
\end{figure}
%%%%%%%%%%%%%%%%%%%%%%%%%%%%%

\section{Summary and outlook}\label{sec:summary}

We investigated structures of low-lying $(0s)_\Lambda$ states 
 in $p$-shell  $\Lambda$ hypernuclei with microscopic cluster models. 
To describe structures of the ground and excited states of core nuclei, 
we applied the GCM of microscopic $\alpha+d$, $2\alpha$, and $2\alpha+d$ cluster models for $^{6}\textrm{Li}$,
$^{8}\textrm{Be}$, and $^{10}\textrm{B}$, respectively, 
and that of $2\alpha+t(h)+p_{3/2}$ and $3\alpha+p_{3/2}$ models with the cluster breaking
for $^{11}\textrm{B(C)}$ and $^{12}\textrm{C}$. The
$0s$-orbit $\Lambda$ particle in $\Lambda$ hypernuclei is treated by 
the single $S$-wave channel calculation with the $\Lambda$-nucleus potentials, which are  
constructed by folding the effective $\Lambda$-$N$ interactions with the nuclear density obtained 
by the microscopic cluster models. As a core polarization effect, the core size reduction is taken into account in a simple way.

For $A>10$ systems, the core polarization, i.e., the nuclear size reduction 
by the $\Lambda$ particle is small. 
The small change of the core size in the ground state of  
$^{13}_\Lambda\textrm{C}$ is consistent with
prediction of other calculations. Moreover, 
the core polarization effect on energy is minor and regarded 
as a higher-order perturbation in the $\Lambda$ binding except for $A<10$ systems.

We discussed energy spectra of the ground and low-lying excited states. 
A particular attention is paid to excitation energy shift and its 
relation to nuclear size difference between the ground and excited states in each $^{A}_\Lambda{Z}$ system.
The present results show a  clear correlation between the energy shift and the size difference.
Namely, the larger size difference, the larger excitation energy shift. 
%The calculated and experimental energy shifts are plotted against the calculated size difference. 
The calculation with ESC08a(DI) qualitatively 
describes the systematic trend of the available experimental data for the energy shift. 
The mechanism of this correlation is understood as
the higher nuclear density gives the larger attraction, i.e., the
deeper $\Lambda$-nucleus potential. 
In other words, the $0s$-orbit $\Lambda$ particle can
probe the inner density difference through the 
$\Lambda$-$N$ interactions. 

We also discussed the shrinkage properties of the $\alpha+d$ and $2\alpha$ cluster structures in
$^7_\Lambda \textrm{Li}$ and $^9_\Lambda \textrm{Be}$, respectively, and compare the present results with other 
calculations. The obtained results are similar to those of other calculations.
For $^7_\Lambda \textrm{Li}$,
the results show the significant shrinkage, and  reproduce the experimental
$E2$ transition strengths in $^6\textrm{Li}$ and $^7_\Lambda \textrm{Li}$ without using effective charges. 
Also in $^9_\Lambda \textrm{Be}$,  the significant shrinkage occurs
consistently with other calculations.
%the obtained reduction factor $S$ and $S_{E2}$ are consistent with those of other calculations. 
%the absolute values of the distance $\overline{r}_{\alpha\textrm{-}d}$ are  $10-20$\%  larger 
%the $B(E2)$ value is $40-70$\% larger
%than those of the semi-microscopic cluster model calculations. The present results reproduce the experimental
%$E2$ transition strengths in $^6\textrm{Li}$ and $^7_\Lambda \textrm{Li}$ without using effective charges. 

The effective $\Lambda$-$N$ interactions used in the present calculation are the spin-independent central 
interactions of the $\Lambda NG$ interactions, which were derived from the $\Lambda$-$N$
 interactions of the one-boson-exchange model
based on the $G$-matrix calculation for an infinite nuclear matter. 
We adopted two treatments of the $k_f$ parameter in the $\Lambda NG$ interactions.

One is the density-independent (state-independent) $\Lambda NG$ interactions and the other 
is the density-dependent (state-dependent) $\Lambda NG$ interactions with the averaged density approximation.
The present results indicate that 
the density-independent
ESC08a interactions are rather favored in description of  the systematic trend of experimental 
excitation energy shift  in $^A_\Lambda Z$ than the density-dependent ones.
However, as for the quantitative reproduction, 
the density-independent calculation generally overestimates the experimental energy shift by a factor of $1.5-2$. 
It is likely that weak density dependence of the $\Lambda NG$ interactions 
may be suitable for detailed description of the excitation energy shift  in $^A_\Lambda Z$.
The density-dependent $\Lambda NG$ interactions were constructed based on the $G$-matrix theory 
in an infinite nuclear matter and originally designed to 
reproduce systematics of $\Lambda$ binding energy in $^A_\Lambda Z$ in a wide mass-number region. 
The origin of the density dependence is the Pauli blocking effect on 
intermediate states in $\Lambda$-$N$ scattering processes.  
The Pauli suppression of the effective $\Lambda$-$N$ interactions is stronger in the higher nuclear density.  However, it is not obvious 
that the density dependence of the 
$\Lambda NG$ interactions can properly probe
the density (or structure) difference between 
the ground and excited states
in each $^A_\Lambda Z$. 
The present $k_f$ dependence in the $\Lambda NG$ interactions is likely to be 
too strong to simulate the structure dependence of the effective interactions in low-lying states 
in each system.

The present framework is based on local density approximations in treatment of nuclear density matrices of 
the exchange folding potentials and  that of the density dependence of the $\Lambda$-$N$ interactions. 
The applicability of the present treatments to highly excited (particle-hole) states should be checked. In particular, the applicability to 
very dilute systems with much lower density than the saturation density should be carefully tested.

The present treatments of the $\Lambda$ particle and the core polarization 
are very simple. 
However, one of the 
great advantages is that the method is handy 
and economical, and be able to be applied to general nuclear structure models 
without changing computational codes for the nuclear structure calculation. 
Moreover, the method can be applied to double $\Lambda$ hypernuclei straightforwardly.
%The method might be helpful for systematic study of various hypernuclei. 

In the present work, we used  the spin-independent central $\Lambda$-$N$ interactions and ignored
the $\Lambda$ spin coupling with the core nuclear spin. Within the present framework, 
it is able to discuss only the leading properties of energy spectra. In order to discuss 
detailed energy spectra 
and spin dependence of the $\Lambda$-$N$ interactions, 
some extensions of the framework are needed. 

%\section*{Acknowledgments} 
\begin{acknowledgments}
The author thanks to Dr.~Motoba and Dr.~Isaka for fruitful discussions.
This work was inspired by the  Karuizawa workshop (June 2017).
% organized by Kohno, 
%Nakamoto, Suzuki, and the author. 
The author would like to give a huge thanks to Dr.~Fujiwara for his continuous  
encouragement. 
%She also thanks to Dr.~Akaishi for many discussions 
%on hypernuclei and effective interactions during the time they were working in KEK. 
The computational calculations of this work were performed by using the
supercomputer in the Yukawa Institute for theoretical physics, Kyoto University. This work was supported by 
JSPS KAKENHI Grant Number 26400270.
\end{acknowledgments}

\end{document}